**Sociodemographic inequalities in student achievement:**

**An intersectional multilevel analysis of individual heterogeneity and discriminatory accuracy**

**(MAIHDA) with application to students in London, England**

Lucy Prior[1*], Clare Evans[2], Juan Merlo[3] and George Leckie[1]

[1] Centre for Multilevel Modelling and School of Education, University of Bristol, 35 Berkeley Square, Bristol, BS8 1JA, United Kingdom

[2] Department of Sociology, 1291 University of Oregon, Eugene, OR 97403, USA

[3] Unit for Social Epidemiology, Department of Clinical Sciences in Malmö, Lund University, Sweden

*Corresponding author: Email: lucy.prior@bristol.ac.uk, Phone: +44 (0)117 3314633, Address: School of Education, 35 Berkeley Square, Bristol, BS8 1JA, United Kingdom

**Funding**

This work was funded by United Kingdom Economic and Social Research Council (ESRC) grants ES/R010285/1 and ES/W000555/1.

**Acknowledgements**

This work was produced using statistical data from the United Kingdom Office for National Statistics (ONS). The use of the ONS statistical data in this work does not imply the endorsement of the ONS in relation to the interpretation of analysis of the statistical data. This work uses research datasets which may not exactly reproduce National Statistics aggregates.

**Keywords**

student achievement; sociodemographic inequalities; intersectionality; multilevel models; interaction effects

**Abstract**

Sociodemographic inequalities in student achievement are a persistent concern for education systems and are increasingly recognized to be intersectional. Intersectionality considers the multidimensional nature of disadvantage, appreciating the interlocking social determinants which shape individual experience. Intersectional multilevel analysis of individual heterogeneity and discriminatory accuracy (MAIHDA) is a new approach developed in population health but with limited application in educational research. In this study, we introduce and apply this approach to study sociodemographic inequalities in student achievement across two cohorts of students in London, England. We define 144 intersectional strata arising from combinations of student age, gender, free school meal status, special educational needs, and ethnicity. We find substantial strata-level variation in achievement composed primarily by additive rather than interactive effects with results stubbornly consistent across the cohorts. We conclude that policymakers should pay greater attention to multiply marginalized students and intersectional MAIHDA provides a useful approach to study their experiences.

**Keywords:**

Sociodemographic inequalities in student achievement:

An intersectional multilevel analysis of individual heterogeneity and discriminatory accuracy

(MAIHDA) with application to students in London, England

## 1. Introduction

*Sociodemographic differences in student achievement*

The persistence of substantial sociodemographic differences in student achievement is well-established and a continuing concern for education systems. As with other social inequalities such as those in health, these differentials can be seen as symptomatic of wider social processes, revealing the action of multi-layered social determinants such as discrimination, marginalisation, privilege, and differential access to resources (Ladson-Billings and Tate, 1995; Marmot and Wilkinson, 2005). These social processes act in a myriad of contexts, including schools, families and neighborhoods (Braveman and Gottlieb, 2014). Researchers typically rely on particular social identities to proxy the action of these processes, for instance the use of ethnicity to represent the action of structural racism. In evaluating student inequalities, studies and official statistics have primarily focused on achievement gaps between more and less socioeconomically advantaged students (Chmielewski, 2019; Hunt et al., 2022), across ethnicities (Reardon, Kalogrides and Shores, 2019, Bradley et al., 2007, Leckie and Goldstein, 2019), by gender (Qualifications and Curriculum Authority, 2008), or by student special education needs status (Education Endowment Foundation, 2021).

*Intersectionality*

A longstanding weakness in many studies of sociodemographic inequalities in student outcomes is that by focusing on singular sociodemographic factors they ignore how systems of advantage and disadvantage are inherently interlocking and entangled. In doing so, this can risk masking important heterogeneity across the concurrent experience of different combinations of categories of other sociodemographic factors (Hernández-Yumar et al., 2018).

In response to this, researchers have called for increased attention to intersectionality. Originally articulated by Crenshaw (1989) and Collins (1990/2009), intersectionality is a critical Black feminist theory that confronts the interlocking, mutually constituted nature of systems of oppression, including racism, sexism, and classism. These systems create positionalities at the intersection of multiple systems of marginalization. For example, the position of Black women, who often experience harms above and beyond what might be expected from unidimensional understandings of being Black or a woman. Social systems of discrimination, marginalization and oppression are thus viewed as shaping individual life chances and experiences through their multi-dimensional social identities (Bell, Holman and Jones, 2019; Green, Evans and Subramanian, 2017; Homan, Brown and King, 2021).

In recent years, the intersectional framework has expanded to consider the uniqueness of intersectional identities beyond ethnicity and gender, including those that experience social privilege. Researchers have also applied a range of quantitative methods to explore intersectionality, including linear regression, classification and regression trees, conditional inference trees, random forests, and multilevel intersectional regression (Bauer, 2014; Evans et al., 2018; McCall, 2005, Mahendran, Lizotte and Bauer, 2022). Intersectional "theory" can therefore be thought of as a framework for analysis, which presupposes the uniqueness of social experience and uses that framing for making the invisible visible and critiquing the power structures underlying the social production of inequalities.

Studying multiple interlocking social positionalities may therefore help to inform a richer understanding of educational and other inequalities (Jones, Johnston and Manley, 2016), through gaining insight into how upstream social determinants interweave to manifest in individual outcomes. There is increasing recognition of the importance of intersectionality in education research, though engagement with intersectionality theory in quantitative settings remains limited for now. For instance, in the US, researchers have recognised the entanglements between ethnicity and socioeconomic status (SES) as important to educational outcomes (Hung et al., 2020; Paschall, Gershoff and Kuhfeld, 2018). In the UK there has been a recent focus on 'left behind' White students from disadvantaged backgrounds (House of Commons Education Committee, 2021). These examples show growing appreciation for how systems of oppression related to racism and the power hierarchies of social status may combine to influence educational outcomes. However, there remains a space for greater attention to the evaluation of multiple, interlocking disadvantages in quantitative educational research.

*Individual heterogeneity*

A second weakness in studies of sociodemographic inequalities in student outcomes is the focus on group mean differences, with often no attention given to student heterogeneity around these group means. In health studies, this concern has been referred to as the 'tyranny of averages' (Merlo et al., 2017). Ignorance of student variation around group means may lead to the attribution of the same 'average' student outcome to all students within a studied group. In the presence of substantial student outcome heterogeneity, this could lead to unnecessary stigmatization of higher performing students in lower mean performing groups and false expectations for many lower performing students in higher mean performing groups. Similarly, interventions aimed at improving student achievement and other outcomes may be inefficient if they are targeted only based on the information provided by group means (Merlo, Wagner and Leckie, 2019). Where there is a large degree of student outcome heterogeneity, targeted interventions are likely to miss students who are struggling but not captured by the disadvantaged group identifier defining group averages. Using information on both group means and the distribution of individual values around those groups means, especially in relation to other student characteristics, is necessary to avoid these negative consequences.

*Policy implications for intersectionality and individual heterogeneity*

Identifying and understanding intersectional effects on student achievement has important implications for educational policy. For example, in England some have called for targeted interventions and support for Black boys, as an intersection of sociodemographic characteristics showing especially low mean achievement and overrepresented in exclusions (Byfield and Talburt, 2020; Wint et al., 2022). In contrast, 'Pupil Premium' funding in England, which gives higher per student funding for certain children, is focused only on socioeconomic disadvantage, primarily measured through free school meal status (Department for Education, 2022a). However, the action of intersectional effects may make it more efficient to consider multiple intersecting sociodemographic characteristics that may mitigate or exacerbate the impacts of socioeconomic disadvantage.

In order to understand whether universal or targeted policies would be more effective it is also important to appreciate student outcome heterogeneity around group means. For instance, a study may identify a group mean difference where boys score, on average, 10 points lower on an achievement test than girls. Based on this information, policymakers may choose to enact policies targeted at boys to raise their mean achievement. However, these types of targeted interventions may only be efficient where there is little overlap in the distribution of scores around the girl and boy mean scores, or in other words where student outcome heterogeneity is small. In contrast, when there is substantial overlap in the score distributions, with many boys scoring higher than the mean girl and many girls scoring lower than the mean boy, a universal approach to raising academic achievement, directed at all students, may be more efficient. However, a more nuanced third approach, whereby an overall goal (e.g., a benchmark of educational achievement) is facilitated by a universal intervention that is proportionally tailored to meet the specific needs of the different groups (e.g., greater scale and intensity for the most disadvantaged groups) may prove most effective (Carey, Crammond and De Leeuw, 2015; Marmot and Bell, 2012; Merlo et al., 2019). Therefore, researchers and policymakers must simultaneously consider intersectionality and individual heterogeneity as two interlinked phenomena in their attempts to quantify, comprehend, and tackle educational inequalities.

*Aim and outline of article*

Multilevel analysis of individual heterogeneity and discriminatory accuracy (MAIHDA) is a promising new approach to quantitative intersectional analysis which simultaneously considers intersectionality and individual heterogeneity. However, to date, it has almost exclusively been applied in the study of individual outcomes in population health, for instance in investigations of substance abuse (Permark et al., 2019), chronic obstructive pulmonary disease (Axelsson Fisk et al., 2018), and body mass index (Hermández-Yumar et al., 2018) with only one application to educational research (Keller et al., 2022). The aim of this article is to introduce intersectional MAIHDA as an approach to study and understand student achievement and to illustrate this approach with an application to student achievement in London, England. Specifically, we compare and contrast intersectional variation and specific effects in two cohorts of students at different stages of the educational life course, the end of primary schooling (age 11) and the end of compulsory secondary schooling (age 16) and in doing so describe the persistence of intersectional effects over time. The article proceeds with a description of the methods and data, followed by the results. We finish with a discussion where we reflect on the implications of our findings for educational systems and the pursuit of educational equality, as well as our reflections on intersectional MAIHDA as an approach to study intersectionality.

## 2. Intersectional MAIHDA

*Overview*

Recently, several research groups have developed a multilevel linear regression approach to quantitatively assess intersectional variation (Evans, 2015; Evans et al., 2018; Evans, Leckie and Merlo, 2020; Merlo, 2018; Jones et al., 2016): multilevel analysis of individual heterogeneity and discriminatory accuracy, referred to as intersectional MAIHDA. This development was made in response to critiques of the conventional linear regression approach, whereby intersectional effects are assessed through the inclusion of interaction terms between two or more aspects of social identity.

Intersectional MAIHDA involves treating intersecting social positions as distinct contexts within the multilevel framework in much the same way that schools are typically applied as contextual units when studying student outcomes (Raudenbush and Bryk, 2002). Individuals with certain identities and resources thus occupy particular intersectional positionalities within interlocking hierarchies, and they may share social experiences (e.g., marginalization, discrimination, advantages) as a consequence. This approach is argued to align with the tenets of intersectional theory in that the social categories within which individuals belong are treated as a feature of their intersectional stratum, rather than of the individuals themselves (Evans et al., 2018). This helps to maintain an appreciation for how these social categories, and any differentials in mean outcomes across strata, are symptoms of wider social processes and structural power hierarchies (Evans et al., 2020). This helps avoid 'blaming the victim' in perspectives on inequalities (Persmark et al., 2019). Additionally, treating intersectional strata as contexts helps circumvent issues around the tendency to implicitly reinforce conceptualisations of certain groups (White, male, etc.) as the 'norm' through the choice of these aspects as 'reference' categories (Evans et al., 2018; Persmark et al., 2019).

Intersectional MAIHDA is argued to have several advantages over the conventional approach. First, increased scalability in that it is simple to include many intersections across multiple social dimensions simultaneously (Hernández-Yumar et al., 2018). In contrast, the inclusion of many higher-order interactions can become interpretationally and methodologically challenging in the conventional approach (Bell et al., 2019). Second, the multilevel approach may be viewed as more parsimonious as the distribution of the intersectional stratum random effects are summarised through a single parameter, the random effects variance. The random effects are predicted postestimation in contrast to the conventional approach where all interaction effects are estimated directly via regression coefficients (Evans et al., 2020). Third, in the multilevel approach, the predicted intersectional stratum effects are empirical Bayes predictions, whereby the predicted values are shrunk towards the overall average. This reduces the risk of overinterpreting extreme predictions that may have arisen by chance (Jones, et al., 2016; Persmark et al., 2019) and so addresses concerns around multiple testing (Bell et al., 2019). Fourth, through the multilevel approach, researchers can evaluate variability within and between strata, aligning with key themes of complexity and heterogeneity from intersectional theory (Evans et al., 2018; 2020; Hernández-Yumar et al., 2018; Homan et al, 2021). Fifth, in contrast to the conventional approach, MAIHDA provides direct quantification of the discriminatory accuracy of the information provided by intersectional strata to classify individuals in terms of their outcomes. This type of information is then helpful to decision making around the use of targeted, universal, or proportionally targeted interventions to improve outcomes. Specifically, where between-stratum variation is relatively high this relates to individuals being more homogeneous within particular intersectional contexts and thus tailored policies aimed at specific strata could be more effective.

The intersectional MAIHDA approach consists of fitting two multilevel linear regression models. We describe each model below in the context of our application to student achievement where students (level 1) are viewed as nested in intersectional strata (level 2).

*Model 1: Unadjusted model*

The first model, Model 1, is a random-intercept linear regression with no covariates. The purpose of this unadjusted model (null or simple intersectional model) is to estimate the extent to which intersectional strata explain overall variation in student achievement outcomes and to quantify the total degree of inequality between strata. Let $y_{ij}$ denote the achievement score of student $i$ ($i = 1, ..., n_j$) at level-1 in the data hierarchy in strata $j$ ($j = 1, ..., J$) at level-2 in the data hierarchy. Model 1 can then be written as

$$y_{ij} = \beta_0 + u_j + e_{ij} \qquad [1]$$

where $\beta_0$ is the intercept measuring the average of the observed stratum mean achievement scores, $u_j$ denotes the random intercept effect associated with stratum $j$ measuring how much higher or lower mean achievement is in that stratum relative to the overall mean, and $e_{ij}$ denotes the student-level residuals measuring how each student's individual score deviates from their stratum mean. The intercept is estimated as a precision-weighted average of the observed strata means where the weights account for not just stratum size but the overall degree of stratum clustering in student outcomes (Evans et al., 2020, Raudenbush and Bryk, 2002). Thus, while higher weight is always given to larger strata, the strength of this relationship weakens with increased clustering. The stratum random effects and student residuals are modelled as normally distributed with zero means, and between- and within-stratum variances of $\sigma_u^2$ and $\sigma_e^2$ respectively.

$$u_j \sim N(0, \sigma_u^2) \qquad [2]$$

$$e_{ij} \sim N(0, \sigma_e^2) \qquad [3]$$

The variance partitioning coefficient (VPC) can then be calculated, measuring the proportion of overall variation in student achievement that is attributable to intersectional strata. Thus, the VPC is a general measure of how well different social intersections can distinguish student achievement outcomes, that is a quantification of the degree of overlap in the stratum-specific outcome distributions. The VPC is calculated as

$$\text{VPC} = \frac{\sigma_u^2}{\sigma_u^2 + \sigma_e^2} \qquad [4]$$

Postestimation, empirical Bayes prediction can be used to assign values to the stratum random effects and 95% confidence intervals. The summations of the estimated intercept and the predicted random effects, $\beta_0 + u_j$, then give predicted stratum means which can be studied to identify specific combinations of student characteristics associated with higher and lower student achievement. These predictions differ and are preferred to observed stratum means in that they exhibit shrinkage (smoothing) whereby the values are shrunk towards the overall average in inverse proportion to stratum size to protect against overinterpreting extreme predictions that may have arisen by chance (sampling variation).

*Model 2: Adjusted model*

The second model, Model 2, extends Model 1 by adding the categorical sociodemographic characteristics used in construction of the intersectional strata as main effects covariates (e.g., age, gender, FSM, SEN and ethnicity). The purpose of this adjusted model (main effects or interaction model) is to assess the relative importance of main effects versus two- or higher-way interaction effects between the characteristics in explaining the intersectional stratum variation in student achievement established in Model 1. Where interaction effects are found, this indicates that the unique social experiences of individuals at that intersection have produced an impact (advantageous or disadvantageous) different from that implied by the sum of the separate singular main effect influences of the studied characteristics. Note, however, that where no interaction effects are found, this does not invalidate intersectional thinking – it merely suggests that for that particular outcome, in that particular population, additivity describes patterns of inequality reasonably well.

Model 2 is written as

$$y_{ij} = \mathbf{x}_j'\boldsymbol{\beta} + u_j + e_{ij} \qquad [5]$$

where $\mathbf{x}_j$ denotes the vector of the intercept and stratum-level covariate values for the categorical components of the intersections entered as dummy variables for stratum $j$. The vector of regression coefficients measuring the main effects of these covariates is then denoted by $\boldsymbol{\beta}$, whereas the stratum random effects $u_j$ now capture all two-and higher-way interaction effects between these variables. Specifically, the stratum random effects capture the deviation of the mean achievement score for each stratum from that predicted by the additive sum of the effects of the stratum-level covariates. Thus positive (negative) values of $u_j$ correspond to strata whose mean achievement scores are higher (lower) than that we predict or expect from the additive sum of effects. The variance attributable to these interaction effects is measured by $\sigma_u^2$.

The relative importance of the main effects in explaining the overall intersectional variation can be assessed via the proportional change in the stratum variance (PCV) statistic associated with moving from model 1 to 2. We can write this as

$$\text{PCV} = \frac{\sigma_{u(1)}^2 - \sigma_{u(2)}^2}{\sigma_{u(1)}^2} \quad [6]$$

where $\sigma_{u(1)}^2$ and $\sigma_{u(2)}^2$ represent the stratum-level variance from model 1 and model 2 respectively.

Postestimation, we can again predict and study the stratum random effects. These effects now allow us to identify, among all possible combinations of sociodemographic characteristics, those combinations with the most positive or negative interaction effects. That is, those strata whose mean achievement deviates most from that predicted by the additive effects of their characteristics.

## 3. Data

This study uses data drawn from the Department for Education (DfE) National Pupil Database (NPD), an administrative dataset of all state-educated students in England. We focus on two cohorts of students, at age 11 and age 16 in the 2018/19 academic year attending mainstream, state-funded schools in London. Our analysis samples consist of 90,744 students in the age 11 cohort, and 71,321 students in the age 16 cohort.

Our outcome is student achievement, as measured by student performance in English and mathematics tests at the end of primary school (year 6, age 11, Key Stage 2 tests) and English, mathematics and six further subject examinations at the end of secondary school (year 11, age 16, General Certificate of Secondary Education examinations). To facilitate interpretation, we analyse standardized versions of the resulting age 11 and age 16 achievement scores (means of 0 and SDs of 1) (Supplementary Figure S1).

We consider a set of student sociodemographic characteristics to represent intersectional positionalities (Table 1). These are student term of birth (autumn: September, October, November, December; spring: January, February, March, April; summer: May, June, July, August), gender (boy, girl), whether the student has been eligible for free school meals (FSM) in the previous six years (no-FSM, FSM), special educational needs (SEN) status (no-SEN, SEN), and ethnicity (White, Black, Asian, Mixed, Other, Unclassified). This set of characteristics represent aspects of student experience known to relate to student achievement (Leckie and Goldstein, 2019; Navarro, García-Rubio and Olivares, 2015) and as a result are monitored annually in England (Department for Education, 2022b). Through the lens of intersectional theory, these characteristics act as proxies for wider, upstream social determinants (student gender for instance marking any educational influences from gendered power hierarchies). We then define intersectional strata as the set of all possible combination of these five characteristics giving 144 intersections in total ($3 \times 2 \times 2 \times 2 \times 6$). The number of students per stratum ranged from 11 to 4,052 at age 11, and 11 to 3,201 at age 16 (Table S1) There are a small number of strata with fewer than 10 students which are included in the modelling but excluded from the results and figures to avoid disclosure.

Analysis is conducted using Stata version 17 (StataCorp, 2021). An example Stata code file and simulated data to replicate an intersectional MAIHDA analysis is provided with the supplementary materials.

## 4. Results

*Descriptive summary*

Table 1 presents average age 11 and age 16 scores across the different sociodemographic component identities used to construct our intersectional strata. Focusing first on age 11 scores, we see average achievement gaps that replicate earlier findings in the literature: summer born students on average score lower than autumn born students (0.22 SD difference); girls score on average higher than boys (0.14 SD difference); those eligible for FSM score on average much lower than those not eligible for FSM (0.48 SD difference); SEN students score on average substantially lower than those without SEN (1.14 SD difference); and Asian students tends to have higher achievement (0.18 SD difference) and Black students lower (-0.20 SD difference) than White students. The overall picture is similar for age 16 students but with some gaps wider (gender, FSM, Asian-White) and others narrower (age, SEN). These findings agree with those previously reported (Commission on Race and Ethnic Disparities, 2021; Crawford, Dearden and Meghir, 2010; Prior, Goldstein and Leckie, 2021). Summaries of the average age 11 and age 16 achievement scores for each intersectional stratum, as well as the proportion of students falling into each category combination, are presented in Supplementary Table S1.

*Model 1: Unadjusted model results*

Table 2 presents the Model 1 results predicting age 11 and age 16 student achievement. These intercepts are estimated as -0.402 SD and -0.372 SD at age 11 and age 16. That both estimates are substantially below the sample mean student achievement scores of 0 reflects the fact that the intercept is a precision-weighted rather than stratum-sized weighted average of the stratum means, the degree of clustering is substantial (see below), and smaller strata tend to be substantially lower scoring than larger strata (see Supplementary Figure S2).

In Model 1, the VPC statistics show a large proportion of the overall variance in student achievement positioned between intersectional strata at age 11 (around 30%), and a slightly smaller though still substantial amount at age 16 (25%). In comparison, we typically see between 10 to 20% of variance in student achievement positioned at the higher level when nesting students within schools (Reynolds et al., 2014). This suggests that the intersectional stratum in which a student belongs is a significant factor in their associated academic achievement, both at age 11 and age 16, with a large degree of discriminatory accuracy by which to classify student outcomes. The high VPC suggests policies tailored to specific intersectional groups may be effective or beneficial for closing educational gaps. Intersectional strata could thus be an equally if not more important context than schools to consider for interventions to improve student outcomes. Future work could employ a cross-classified structure of students nested within both their schools and intersectional strata to explore this further.

The Model 1 predicted stratum means ($\beta_0 + u_j$, for each stratum $j$) provide estimates of the mean achievement score in each stratum. These predictions are therefore strongly related to the observed stratum means (Supplementary Figure S3), the difference being the shrinkage applied. As can be seen in Supplementary Figure S4, which plots the difference between the shrunken and un-shrunken stratum effects for Model 1 against stratum size, it is small strata with extreme stratum effects which receive the highest degree of shrinkage. For example, the largest difference between shrunken (-0.84) and un-shrunken (-0.72) stratum effects at age 11 is shown by a stratum with just 15 students. This demonstrates the benefit of using the multilevel model and predicted stratum means over observed stratum means: without the application of shrinkage researchers' risk overinterpretation of extreme and unreliable stratum means.

Table 3 presents the top and bottom 10 ranked strata in terms of the predicted stratum means $(\beta_0 + u_j)$. The main effects of the sociodemographics dominate with the patterns reflecting the known average achievement gaps noted in the descriptive statistics. The bottommost strata at age 11 are typically characterised by FSM eligible and SEN identities, whilst the topmost show more heavy representation for non-FSM, non-SEN, older in year students, and girls. Some of these patterns seen in Table 3 are more strongly evident at age 16 than age 11, particularly in relation to gender and this reflects the larger overall gender gap at age 16 compared to 11. The most extreme strata show various ethnicities, however, Asian ethnicities do appear in the best performing strata at both age 11 and age 16.

The top row of Figure 1 presents the predicted stratum effects ($u_j$, which represent how much each stratum deviates from the overall mean – the intercept) in rank order together with their 95% confidence intervals for age 11 and age 16 achievement. The vast majority of these stratum effects are significantly different from average, with 94% of effects at age 11 significantly above or below average and 81% at age 16. This reflects the high VPCs and the generally large stratum sizes at each age. The plot for age 11 shows a particularly strong distinction between those stratum random effects which are below and above average. As a result, the stratum effects show a bimodal distribution, rather than being normally distributed (see Supplementary Figure S5). Examination of the stratum components reveals that this is primarily driven by a wide unadjusted SEN gap whereby the lowest ranked strata involve those with SEN, and the higher ranked strata those without SEN (and to a lesser extent similar patterning for the FSM and non-FSM groups).

Additionally, in Figure 1 we have added dashed lines as representations of aspirational benchmarks of high achievement. These correspond to achieving a 'high standard' (Department for Education, 2022c) at age 11 (strata effect of 0.94 SD) and gaining eight Grade A's in the age 16 examinations (0.59 SD). These standards could act as universal goals for school systems, and the presentation of the strata random effects against these shows which strata may be suffering impediments to their achievement and thus require tailored interventions to meet the overall benchmark. For these example benchmarks, 5% of strata meet and exceed the age 11 benchmark, whilst 17% meet or surpass the age 16 target.

Figure 2, top row, shows the predicted stratum effects and ranks are very stable across age 11 and age 16 (Pearson and Spearman rank correlations around 0.90). Nevertheless, the relative standing of some strata changes considerably between age 11 and age 16. The biggest negative differences (where the age 16 cohort typically scores lower than the age 11 cohort) are seen by the strata typically defined by being a boy, FSM eligible, of an Unclassified ethnicity and without SEN. The largest difference (-0.48 SD) is seen for this group combined with being Summer born. These most negative differences could thus relate to a tendency for socioeconomically disadvantaged boys to fall behind in relation to other groups over the course of secondary schooling. In contrast, the strata with the largest positive differences (indicating the age 16 cohort tends to score higher than the age 11 cohort for that particular set of sociodemographic characteristics) were seen for the two strata defined by being Autumn born, a girl, with SEN, and of Asian ethnicity. The largest shift (0.53 SD) was seen for these characteristics combined with not being eligible for FSM, whilst the second largest difference (0.44 SD) saw this combination with eligibility for FSM. Therefore, in the case of increasing stratum scores between the age 11 and age 16 cohorts, older in year, Asian girls are gaining progress in relation to other groups. This aligns with known trends regarding the performance of female students and Asian minorities throughout secondary education (Leckie and Goldstein, 2019).

*Model 2: Adjusted model results*
Our second model includes the sociodemographic components of the strata as covariates in the models for age 11 and age 16 achievement (Table 2). The results present conditional gaps, as each coefficient represents the effect of that characteristic having considered the other factors. Therefore, the gaps appear smaller than the marginal gaps identified in Table 1. Nevertheless, the patterns of results echoes what we would expect given the descriptive statistics and prior literature (Table 1): younger students score lower (-0.114 SD for Summer born vs. Autumn born) all else equal, girls score higher than boys (though by a small amount, 0.019 SD, and not significant), those eligible for FSM (-0.282 SD) and especially those with SEN (-1.067 SD) are associated with lower scores, and Asian and Mixed ethnicities tend to score higher (0.067 SD and 0.044 SD) than White and especially Black students. Similar relationships are also demonstrated with age 16 achievement, though as with the descriptive analyses, some relationships appear to widen for this older cohort, particularly the gender gap (0.191 SD) and FSM gap (-0.419 SD).

Comparing the variance positioned at stratum level in Model 2 and Model 1 we see a PCV of approximately 97% at age 11 and 96% at age 16. This shows that for these cohorts of students in London, additivity – that is an accumulation of the conditional effects of the sociodemographic components – describes the intersectional variation reasonably well. The remaining stratum-level variance represents variation in achievement resulting from two- or higher-way interaction effects of these components.

The VPC statistics now quantify the proportion of variation in student achievement beyond the main effects that is attributable to stratum interaction effects as opposed to within stratum variability. These VPC statistics are very low (1.3% at both age 11 and age 16) suggesting these interaction effects are very small and that other factors beyond those considered here dominate in explaining the residual variability.

Examination of the bottom-row of caterpillar plots in Figure 1 further demonstrates that the main effects of the stratum components are powerful explanators of intersectional variation: very little variation remains between strata. For Model 2 the predicted stratum random effects are representations of interaction effects, or deviations from additivity. Positive (or negative) effects correspond to strata whose predicted mean scores exceed (or subceed) that predicted by the additive main effects. The predicted stratum random effects for both age 11 and age 16 are small in size. The range in predicted stratum random effects at age 11 and age 16 are both roughly 0.4 SD of student achievement for Model 2, compared with ranges of around 2.0 SD for Model 1. However, there are still many predicted stratum random effects which are significantly different from average (around 38% of strata at age 11 and 36% at age 16). Additionally, 34 stratum effects (24%) are significantly different from average and larger than 0.1 SD in absolute magnitude for age 11 achievement (32 strata, 22% for age 16 achievement). In educational research an effect size of 0.1 SD would often be considered substantively meaningful (Kraft, 2020).

Presentation of the top and bottom 10 predicted stratum effects from the adjusted model for the two cohorts are shown in Table 4. These show that these most extreme stratum effects no longer reflect the associated patterns of the main effects as they have been controlled in the fixed part of the model. For instance, groups with no SEN and not eligible for FSM are now present in the bottom 10 strata and there is a mix of boys and girls in groups comprising the top and bottom strata for both cohorts. Figure 2 shows that the stratum interaction effects are also relatively stable between the age 11 and age 16 cohorts with correlations around 0.70. Thus, in general strata whose mean scores at age 11 are above those implied by simply adding the main effects, tend to also show above predicted mean scores for the age 16 cohort. Nevertheless, despite these relatively high correlations there is still a considerable degree of change in the stratum random effects of different intersections, as evident from the crossovers in Supplementary Figure S6. The largest differences between age 11 and age 16 stratum interaction effects are in the absolute magnitude of 0.17 SD.

The Model 2 predicted stratum random effects represent two- and higher-way interactions between the sociodemographic components. We focus here on potential interactions between ethnicity and FSM. Figure 3 presents the Model 2 predicted stratum random effects for age 11 (left column) and age 16 (right column) for the six groups of strata (rows) relating to combinations of the Black, Mixed and White ethnicities for FSM and Non-FSM students. We omit the other ethnic groups for ease of illustration and because they show less extreme effects. Within each row we plot the 12 stratum effects relating to crossing our chosen ethnicity and FSM combinations with the three remaining sociodemographic components: term of birth, gender and SEN.

Figure 3 shows that the average achievement for White or Mixed ethnicity students who are not eligible for FSM tends to be higher than that implied by summing the main effects of the sociodemographic variables (rows 4 and 6). Conversely, the strata characterised by White or Mixed ethnicity students who are eligible for FSM typically demonstrate residuals which are more negative than expected based on the additive main effects, particularly for strata where the students have no SEN (rows 3 and 5). In contrast, for Black students, the FSM patterning is the other way around with Black FSM students showing, on average, more positive stratum interaction effects than is the case for Black non-FSM students (rows 1 and 2). So the mean achievement for Black FSM students is not as low as that implied by summing the main effects of these variables, but for Black non FSM students it is lower than that implied by summing the main effects, particularly for boys. The patterning by FSM and ethnicity combinations appears to be starker at age 16 than age 11. Supplementary analyses where we add each possible two-way interaction term between the sociodemographic characteristics in isolation (Supplementary Tables S2 and S3) reveal that the ethnicity by FSM interaction does become more important for older students (45% PCV at age 11, 85% PCV at age 16).

Figure 3 also shows indications of different patterning for SEN effects across the six rows. For intersections with Mixed FSM students (row 3), and to a lesser extent those for White FSM students (row 5), the stratum interaction effects show intersections with SEN tending to score higher on average than those without SEN. This could indicate that for students of White or Mixed ethnicity, the support they receive in response to their SEN helps to mitigate the disadvantage posed by their FSM status.

## 5. Discussion

*Substantive findings*

In this article we have introduced the recently proposed intersectional MAIHDA approach to educational research and illustrated it with an application to studying student achievement. We analysed how intersections across age-in-year, gender, SEN, FSM and ethnicity related to student achievement for two student cohorts (age 11 and age 16) in London, England. In our unadjusted models, we identified strong variation in student achievement at the level of our intersectional strata, amounting to 30% for age 11 and 25% for age 16. This represents a substantively important contribution to overall variation in student achievement for students in London, showing that our intersectional strata appear to have reasonably strong discriminatory accuracy for classifying student outcomes. This is substantiated in comparison with Keller et al. (2022) who reported a VPC of 15.9% in their MAIHDA study of reading achievement in Germany. We also observed a considerable number of statistically significant and substantively meaningful intersectional effects, including in the adjusted model where they represent the element of intersectional effects composed by interactions. Together, these results suggest the importance of interlocking social positionalities to achievement outcomes, and that the consideration of intersectionality in educational research and policy may help in the understanding of performance gaps and inequalities.

Our analysis found that this strong intersectional variation was primarily composed of the main, additive effects of the sociodemographic components, a finding echoed by Keller et al. (2002) for reading achievement. The inclusion of the main effects of age, gender, SEN, FSM and ethnicity into the models explained almost all of the stratum level variation at both age 11 (97%) and age 16 (96%). Thus, our results reveal a very small degree of the variation in student achievement that can be attributed to two- and higher-way interaction effects between the variables defining the strata. Therefore, in the context of our analyses on this educational outcome for students in London, the intersectional experiences of students are translating to identifiable inequalities, but this appears to be acting primarily as an additive accumulation of disadvantages, with only a modest degree of interaction effects evident. Importantly, this result does not invalidate the importance of intersectional perspectives for educational achievement. Whilst intersectionality theory features the language of 'interaction' prominently, the term is used conceptually (with a meaning more similar to 'intersection') rather than equating with the statistical terminology of interaction effects (ethnicity × sex × SES) as those which are beyond that expected from a purely additive (ethnicity + sex + SES) perspective (Bauer, 2014).

The overall low VPC in Model 2 for both age 11 and age 16 (1.3%) was also shown to mask some important variation across strata, with several intersections demonstrating significant interaction effects in the adjusted model, as both more positive and more negative than expected given additive effects. Within the small degree of interaction effects identified, intersections between socioeconomic disadvantage and ethnicity appear to be the most important, particularly at age 16. This finding may be important from the position of understanding inequalities in educational outcomes, and for educational policymakers in making practical decisions around support and interventions. However, it is important to recall that whilst we can quantitatively assess which social aspects under consideration may account for the largest degree of variation, theoretically and conceptually the social determinants underlying the studied identities remain fundamentally interlinked and, as such, inseparable (May, 2015).

The strong socially patterned inequalities we find in student achievement do demonstrate the value of the intersectional perspective in calling attention to the multidimensional nature of disadvantage. Without appreciation for this 'multiple' nature, where achievement gaps are studied unidimensionally, researchers may miss critical aspects of inequality. For instance, our results highlight the importance of considering the heterogeneity produced from even small interaction effects. We saw this in relation to Black FSM eligible students: in isolation both characteristics relate to generally lower scores on average, whilst strata combining these two categories (having accounted for these main effects) showed some of the highest average strata effects. That is, black FSM students do not score as low as implied by the sum of their separate negative main effects.

Through separate assessments of student achievement at age 11 and age 16 we gained an appreciation for how intersectional variation differed across these two important stages of the educational life course (end of primary and end of compulsory secondary schooling). We identified high correlations between the predicted stratum random effects, particularly so in our unadjusted model (correlations around 0.90). This indicates the entrenched nature of these educational differentials, and their evidence at age 11 supports arguments in favour of addressing inequalities as early as possible (Heckman, 2011). Additionally, we continued to see a high correlation (of approximately 0.70) in the adjusted model, suggesting persistence over time for the interaction effects of our explored sociodemographics. Therefore, through the intersectional framework we can draw attention to the persistent and pervasive influence of socially structured power hierarchies that serve to reinforce specific educational inequalities across different age groups.

*Policy implications*

Evidence for differentials across intersectional strata, whether or not these arise through additive or interaction effects, means the combination of sociodemographic characteristics to which a student belongs is an important indicator for student outcomes. The high VPC in Model 1 suggests the use of combinations of sociodemographic characteristics could be helpful for identifying hidden subgroups of vulnerable students in terms of their academic performance, and it may be effective to implement targeted policies to these intersectional groups to work towards general aims of equality of achievement. Nevertheless, there remains a high degree of heterogeneity within each intersectional stratum, substantiating the importance of considering individual variation around stratum means, and suggesting it would also be helpful to move beyond sole reliance of mean differences across unidimensional characteristics in the monitoring of educational performance within schools.

The overall importance of intersectional effects suggests it may be important for those within school systems to appreciate how some students may be multiply marginalized, and to pay attention to the performance of social power hierarchies by peers, teachers and through policy within the school setting. For example, in both the US and the UK, Black boys may experience disproportionate disciplinary action for classroom behaviour compared to their peers, resulting in greater exclusions and heightened detriment to academic performance (Demie, 2019, Government Accountability Office, 2018). Furthermore, it may be profitable for educational policy to consider the multidimensionality of social contexts when targeting students for support. For instance, pupil premium funding in England focuses primarily on FSM status to determine the application of funds. However, in our supplementary analyses where we refitted Model 2 with one covariate at a time (Supplementary Tables S4 and S5) we showed that in the case of our two student cohorts, FSM status explained a relatively small degree of variation at age 11 (PCV = 5%) and at age 16 (PCV = 16%) versus the simultaneous consideration of all five sociodemographic characteristics (age 11 PCV = 97%; age 16 PCV = 96%).

These policy implications offer more proximal interventions that may help to address educational differentials. However, it is vital to recall that intersectionality is a critical theory, centred around challenging the systems of power which serve to disadvantage and marginalize certain groups. Therefore, through providing quantitative evidence for intersectional inequalities in student achievement, and thus evidence for the adverse impacts of these oppressive social systems, this study lends weight to calls for wider social reforms to address socioeconomic inequality and structural discrimination.

*Methodological reflections*

This analysis has taken a promising method growing in popularity in health studies and explored its application to educational research, specifically studies of student achievement. In relying on multilevel models, MAIHDA is easy to implement in standard software and the multilevel structure of nesting students within social contexts will be familiar to many educational researchers accustomed to working with the hierarchy of students nested within schools. As with previous applications to health outcomes (Axelsson Fisk et al., 2018, Hermández-Yumar et al., 2018, Permark et al., 2019), we find similar benefits to the multilevel approach. For example, the quantification of the overall importance of the studied intersections via the VPC, understanding of the balance between additive and interaction effects in composing the intersectional variance through the PCV, and allowance for direct examination of predicted stratum means which are more reliable than observed stratum means through the precision weighting of predictions. The multilevel method for exploring intersectional variation has the potential to reveal inequalities in student outcomes in more detail than before.

We would also like to reflect on some limitations to the method. Whilst the approach theoretically allows for numerous variables as inputs to the creation of intersectional strata, here we focus on a limited set of variables (age in year, gender, ethnicity, SEN and FSM). This was primarily driven by what was available in our data, a common issue for researchers. Additionally, as with any quantitative analysis, we necessarily rely on quantifiable categories as proxies for the action of complex social processes (racism, sexism, classism etc.). The multilevel approach can help to identify the tendency for certain strata to perform higher or lower on average, but cannot capture the full complexity of lived experience and social contexts, or provide the underlying reasons behind an identified effect. Qualitative research is needed to understand why intersectional effects arise. The number of variables used in applications will also be feasibly limited by the interpretability of results, particularly where examination of the stratum effects is of interest. The more variables there are the more strata there will be and the greater the challenge in identifying the main patterns in the data. Therefore, theoretically justifiable and important social positionalities may not be included for practical reasons (Hernández-Yumar et al., 2018). We used graphical presentation of the predicted stratum effects in order to convey some of the potential interaction effects (Figure 3), but even here we limited ourselves to a subset of the 144 strata in order to ease interpretability. In order for the results of intersectional analyses such as this to have utility for policymakers, researchers must develop simple methods for disseminating the importance of intersectional effects. Moreover, it remains likely that identified interactional effects could reflect the main effects of various omitted variables not included in the strata creation and so the effects should be viewed as descriptive not causal (Persmark et al., 2019). In this application, we limited our focus to the intersectional strata, however, schools remain an important context for student outcomes. Future studies could incorporate the simultaneous investigation of schools and intersectional groups through cross-classified multilevel models. Further reflections on the methodology arising from this specific application are presented in the supplementary material.

## 6. Conclusion

The intersectional MAIHDA approach applied in this study is a promising approach for the quantification and exploration of intersectional differentials in student achievement and other educational, social and behavioural outcomes. We demonstrate that the experience of multiple positionalities across interlocking social systems is important to students' educational achievement, drawing attention to multidimensionality in the social patterning of inequalities. Further exploration of the multilevel approach to exploring intersectionality across other educational contexts, cohorts and student outcomes has great potential to reveal further insights into the importance of and action of intersectional effects.

**Ethics statement**

Research was proposed and conducted under an Economic and Social Research Council (ESRC) grant which has been granted ethical approval by the ESRC. Informed consent prior to partcipation in the research was not necessary as the research relies on secondary analysis of administrative data collected by the Department for Education in England. Data was shared under a process compliant with data protection legislation and subject to the 5 safes framework.

**Tables**

Table 1.

Descriptive statistics for age 11 and age 16 student achievement and sociodemographic components.

|  |  | Number of students |  | Mean |  | S.D. |
| --- | --- | --- | --- | --- | --- | --- |
| Age 11 score |  | 90,744 |  | 105.42 |  | 8.41 |
| Age 16 score |  | 71,321 |  | 51.09 |  | 19.74 |
| Normalised age 11 score |  | 90,744 |  | 0.00 |  | 1.00 |
| Normalised age 16 score |  | 71,321 |  | 0.00 |  | 1.00 |

|  |  | Number of students |  | Percentage of students |  | Normalised score |  |
| --- | --- | --- | --- | --- | --- | --- | --- |
|  |  | Age 11 | Age 16 | Age 11 | Age 16 | Age 11 | Age 16 |
| Term of Birth | Autumn | 30,358 | 23,591 | 33.45 | 33.08 | 0.11 | 0.05 |
|  | Spring | 29,573 | 23,316 | 32.59 | 32.69 | 0.00 | -0.01 |
|  | Summer | 30,813 | 24,414 | 33.96 | 34.23 | -0.11 | -0.04 |
| Gender | Boy | 45,693 | 35,338 | 50.35 | 49.55 | -0.07 | -0.12 |
|  | Girl | 45,051 | 35,983 | 49.65 | 50.45 | 0.07 | 0.12 |
| Free School Meals | No FSM | 57,985 | 46,500 | 63.90 | 65.20 | 0.17 | 0.19 |
|  | FSM | 32,759 | 24,821 | 36.10 | 34.80 | -0.31 | -0.35 |
| Special Education Needs | No SEN | 74,721 | 61,189 | 82.34 | 85.79 | 0.20 | 0.13 |
|  | SEN | 16,023 | 10,132 | 17.66 | 14.21 | -0.94 | -0.81 |
| Ethnicity | White | 36,739 | 28,070 | 40.49 | 39.36 | 0.01 | -0.03 |
|  | Black | 19,253 | 15,633 | 21.22 | 21.92 | -0.19 | -0.22 |
|  | Asian | 19,288 | 15,434 | 21.26 | 21.64 | 0.19 | 0.28 |
|  | Mixed | 9,636 | 6,898 | 10.62 | 9.67 | -0.02 | -0.03 |
|  | Other | 5,009 | 4,025 | 5.52 | 5.64 | -0.04 | 0.04 |
|  | Unclassified | 819 | 1,261 | 0.90 | 1.77 | -0.02 | -0.02 |

**Table 2.**

Model 1 (unadjusted) and Model 2 (adjusted) results for student age 11 and age 16 achievement.

| | Model 1: Unadjusted | | | | Model 2: Adjusted for main effects | | | |
| | Age 11 score | | Age 16 score | | Age 11 score | | Age 16 score | |
| | Est. | SE | Est. | SE | Est. | SE | Est. | SE |
|---|---|---|---|---|---|---|---|---|
| Intercept | -0.402 | 0.048 | -0.372 | 0.043 | 0.314 | 0.031 | 0.119 | 0.032 |
| Term of Birth | | | | | | | | |
|     Autumn (Ref) | | | | | - | - | - | - |
|     Spring | | | | | -0.066 | 0.025 | -0.041 | 0.025 |
|     Summer | | | | | -0.114 | 0.025 | -0.062 | 0.025 |
| Gender | | | | | | | | |
|     Male (Ref) | | | | | - | - | - | - |
|     Female | | | | | 0.019 | 0.020 | 0.191 | 0.021 |
| Free School Meals | | | | | | | | |
|     No FSM (Ref) | | | | | - | - | - | - |
|     FSM | | | | | -0.282 | 0.020 | -0.419 | 0.021 |
| Special Educational Needs | | | | | | | | |
|     No SEN (Ref) | | | | | - | - | - | - |
|     SEN | | | | | -1.067 | 0.021 | -0.861 | 0.021 |
| Ethnicity | | | | | | | | |
|     White (Ref) | | | | | - | - | - | - |
|     Black | | | | | -0.086 | 0.031 | -0.043 | 0.032 |
|     Asian | | | | | 0.067 | 0.032 | 0.283 | 0.032 |
|     Mixed | | | | | 0.044 | 0.032 | 0.070 | 0.033 |
|     Other | | | | | 0.001 | 0.034 | 0.157 | 0.035 |
|     Unclassified | | | | | -0.005 | 0.045 | 0.037 | 0.043 |
| | Est. | SE | Est. | SE | Est. | SE | Est. | SE |
| Stratum variance | 0.320 | 0.039 | 0.261 | 0.032 | 0.010 | 0.002 | 0.010 | 0.002 |
| Student variance | 0.766 | 0.004 | 0.804 | 0.004 | 0.766 | 0.004 | 0.804 | 0.004 |
| VPC | 29.5% | | 24.5% | | 1.3% | | 1.3% | |
| PCV | - | | - | | 96.8% | | 96.1% | |

Note:

Est. = estimate, SE = standard error. VPC = variance partition coefficient. PCV = proportional change in stratum variance.

**Table 3.**

Model 1 (unadjusted) top 10 and bottom 10 strata means predicting age 11 (top table) and age 16 (bottom table) achievement.

| Model 1: Unadjusted | | | | | | | | | | | | | | | | | |
|---|---|---|---|---|---|---|---|---|---|---|---|---|---|---|---|---|---|
| | | Term of birth | | | Gender | | FSM | | SEN | | Ethnicity | | | | | | Number of students | Strata ID |
| | Strata means | Aut. | Spr. | Sum. | M | F | No FSM | FSM | No SEN | SEN | White | Black. | Asian | Mixed | Other | Unclass. | | |
| **Age 11** — Top 10 strata | | | | | | | | | | | | | | | | | | |
| | **0.57** | ■ | | | | ■ | ■ | | ■ | | | | | ■ | | | 798 | **76** |
| | **0.56** | ■ | | | | ■ | ■ | | ■ | | | | ■ | | | | 2088 | **75** |
| | **0.53** | ■ | | | ■ | | ■ | | ■ | | | | ■ | | | | 2008 | **3** |
| | **0.53** | ■ | | | | ■ | ■ | | ■ | | | | | | | ■ | 81 | **78** |
| | **0.47** | ■ | | | | ■ | ■ | | ■ | | | | ■ | | | | 4022 | **73** |
| | **0.46** | | ■ | | | ■ | ■ | | ■ | | | | ■ | | | | 2143 | **99** |
| | **0.44** | | ■ | | | ■ | ■ | | ■ | | | | | ■ | | | 840 | **100** |
| | **0.41** | ■ | | | ■ | | ■ | | ■ | | | | ■ | | | | 3802 | **1** |
| | **0.40** | ■ | | | | ■ | ■ | | ■ | | | | ■ | | | | 805 | **4** |
| | **0.39** | | ■ | | | ■ | ■ | | ■ | | | | ■ | | | | 1961 | **27** |
| Bottom 10 strata | | | | | | | | | | | | | | | | | | |
| | **-1.09** | | | ■ | | ■ | | ■ | | ■ | | ■ | | | | | 339 | **140** |
| | **-1.11** | | ■ | | | ■ | | ■ | | ■ | ■ | | | | | | 482 | **44** |
| | **-1.11** | | | ■ | ■ | | | ■ | | ■ | ■ | | | | | | 694 | **67** |
| | **-1.12** | ■ | | | | ■ | | ■ | | ■ | ■ | | | | | | 317 | **91** |
| | **-1.12** | ■ | | | | ■ | | ■ | | ■ | | | | | | ■ | 15 | **24** |
| | **-1.13** | | | ■ | | ■ | | ■ | | ■ | | | | ■ | | | 143 | **142** |
| | **-1.14** | | ■ | | | ■ | | ■ | | ■ | | | ■ | | | | 193 | **45** |
| | **-1.14** | | | ■ | | ■ | | ■ | | ■ | ■ | | | | | | 456 | **139** |
| | **-1.23** | ■ | | | | ■ | | ■ | | ■ | | | | ■ | | | 87 | **93** |
| | **-1.27** | ■ | | | | ■ | | ■ | | ■ | | | | | | ■ | 18 | **12** |
| **Age 16** — Top 10 strata | | | | | | | | | | | | | | | | | | |
| | **0.64** | ■ | | | | ■ | ■ | | ■ | | | | ■ | | | | 1666 | **75** |
| | **0.61** | | ■ | | | ■ | ■ | | ■ | | | | ■ | | | | 1600 | **99** |

| | | | | | | | | | | | | | | | | |
|---|---|---|---|---|---|---|---|---|---|---|---|---|---|---|---|---|
| **0.58** | | | | | | | | | | | | | | | 1553 | **123** |
| **0.49** | | | | | | | | | | | | | | | 613 | **76** |
| **0.45** | | | | | | | | | | | | | | | 301 | **77** |
| **0.44** | | | | | | | | | | | | | | | 1616 | **3** |
| **0.43** | | | | | | | | | | | | | | | 126 | **102** |
| **0.42** | | | | | | | | | | | | | | | 672 | **124** |
| **0.42** | | | | | | | | | | | | | | | 3059 | **73** |
| **0.42** | | | | | | | | | | | | | | | 1633 | **51** |

Bottom 10 strata

| | | | | | | | | | | | | | | | | |
|---|---|---|---|---|---|---|---|---|---|---|---|---|---|---|---|---|
| **-1.12** | | | | | | | | | | | | | | | 301 | **44** |
| **-1.13** | | | | | | | | | | | | | | | 210 | **91** |
| **-1.13** | | | | | | | | | | | | | | | 311 | **19** |
| **-1.14** | | | | | | | | | | | | | | | 18 | **48** |
| **-1.15** | | | | | | | | | | | | | | | 126 | **70** |
| **-1.15** | | | | | | | | | | | | | | | 88 | **46** |
| **-1.15** | | | | | | | | | | | | | | | 257 | **20** |
| **-1.16** | | | | | | | | | | | | | | | 242 | **115** |
| **-1.19** | | | | | | | | | | | | | | | 315 | **43** |
| **-1.25** | | | | | | | | | | | | | | | 324 | **67** |

Note:

Aut. = Autumn, Spr. = Spring, Sum. = Summer, M = Male, F = Female, FSM = Free School Meals, SEN = Special Educational Needs, Unclass. = Unclassified. Strata highlighted in bold indicate those with statistically significant effects at the 95% confidence level.

**Table 4.**

Model 2 (adjusted) top 10 and bottom 10 predicted strata means for student age 11 (top table) and age 16 (bottom table) achievement.

Model 2: Adjusted for main effects

| | Strata means | Term of birth | | | Gender | | FSM | | SEN | | Ethnicity | | | | | | Number of students | Strata ID |
|---|---|---|---|---|---|---|---|---|---|---|---|---|---|---|---|---|---|---|
| | | Aut. | Spr. | Sum. | M | F | No FSM | FSM | No SEN | SEN | White | Black | Asian | Mixed | Other | Unclass. | | |
| **Age 11** | **Top 10 strata** | | | | | | | | | | | | | | | | | |
| | **0.18** | ■ | | | | ■ | ■ | | ■ | | | | | ■ | | | 798 | **76** |
| | **0.16** | ■ | | | | ■ | ■ | | ■ | | | | ■ | | | | 2088 | **75** |
| | **0.16** | | | ■ | ■ | | | ■ | | ■ | | | | | ■ | | 116 | **71** |
| | **0.15** | ■ | | | ■ | | ■ | | ■ | | | | ■ | | | | 2008 | **3** |
| | **0.15** | ■ | | | | ■ | ■ | | ■ | | ■ | | | | | | 758 | **7** |
| | **0.13** | ■ | | | | ■ | ■ | | ■ | | ■ | | | | | | 4022 | **73** |
| | **0.13** | | ■ | | | ■ | ■ | | | ■ | ■ | | | | | | 433 | **103** |
| | **0.13** | | | ■ | ■ | | | ■ | | ■ | | ■ | | | | | 606 | **68** |
| | **0.13** | | ■ | | | ■ | ■ | | | ■ | | | | ■ | | | 125 | **118** |
| | **0.13** | ■ | | | | ■ | | ■ | | ■ | | | ■ | | | | 220 | **92** |
| | **Bottom 10 strata** | | | | | | | | | | | | | | | | | |
| | **-0.15** | | ■ | | ■ | | | ■ | ■ | | | | | ■ | | | 424 | **40** |
| | **-0.15** | | ■ | | ■ | | | ■ | ■ | | | | ■ | | | | 957 | **37** |
| | **-0.15** | | | ■ | ■ | | | ■ | ■ | | | | ■ | | | | 565 | **136** |
| | **-0.15** | ■ | | | ■ | | | ■ | ■ | | ■ | | | | | | 1312 | **85** |
| | **-0.16** | ■ | | | ■ | | | ■ | | ■ | | ■ | | | | | 132 | **80** |
| | **-0.17** | ■ | | | ■ | | | ■ | | ■ | | | ■ | | | | 87 | **93** |
| | **-0.17** | | | ■ | ■ | | | ■ | ■ | | ■ | | | | | | 1237 | **133** |
| | **-0.18** | ■ | | | ■ | | | ■ | | ■ | | | | ■ | | | 124 | **81** |
| | **-0.20** | | | ■ | ■ | | | ■ | ■ | | | | | ■ | | | 451 | **64** |
| | **-0.20** | | ■ | | ■ | | | ■ | ■ | | ■ | | | | | | 1207 | **109** |
| **Age 16** | **Top 10 strata** | | | | | | | | | | | | | | | | | |
| | **0.20** | ■ | | | | ■ | ■ | | ■ | | ■ | | | | | | 372 | **79** |

| | | | | | | | | | | | | | | | |
|---|---|---|---|---|---|---|---|---|---|---|---|---|---|---|---|
| **0.15** | | | | | | | | | | | | | | 553 | **31** |
| **0.14** | | | | | | | | | | | | | | 333 | **68** |
| **0.14** | | | | | | | | | | | | | | 383 | **103** |
| **0.14** | | | | | | | | | | | | | | 278 | **89** |
| **0.13** | | | | | | | | | | | | | | 178 | **116** |
| **0.13** | | | | | | | | | | | | | | 231 | **17** |
| 0.12 | | | | | | | | | | | | | | 103 | 34 |
| **0.11** | | | | | | | | | | | | | | 3059 | **73** |
| **0.11** | | | | | | | | | | | | | | 196 | **140** |

Bottom 10 strata

| | | | | | | | | | | | | | | | |
|---|---|---|---|---|---|---|---|---|---|---|---|---|---|---|---|
| -0.13 | | | | | | | | | | | | | | 33 | 11 |
| **-0.14** | | | | | | | | | | | | | | 1084 | **2** |
| **-0.14** | | | | | | | | | | | | | | 111 | **105** |
| **-0.15** | | | | | | | | | | | | | | 941 | **109** |
| **-0.16** | | | | | | | | | | | | | | 826 | **13** |
| **-0.16** | | | | | | | | | | | | | | 1024 | **50** |
| **-0.17** | | | | | | | | | | | | | | 1009 | **133** |
| **-0.17** | | | | | | | | | | | | | | 837 | **37** |
| **-0.17** | | | | | | | | | | | | | | 124 | **129** |
| **-0.20** | | | | | | | | | | | | | | 982 | **85** |

Note:

Aut. = Autumn, Spr. = Spring, Sum. = Summer, M = Male, F = Female, FSM = Free School Meals, SEN = Special Educational Needs, Unclass. = Unclassified. Strata highlighted in bold indicate those with statistically significant effects at the 95% confidence level.

**Figures**

**Figure 1.**

Model 1 (unadjusted, top row) and Model 2 (adjusted, bottom row) caterpillar plots of predicted strata effects for age 11 (left column) and age 16 (right column) achievement. . Dashed lines in top row represent aspirational high achievement benchmarks.

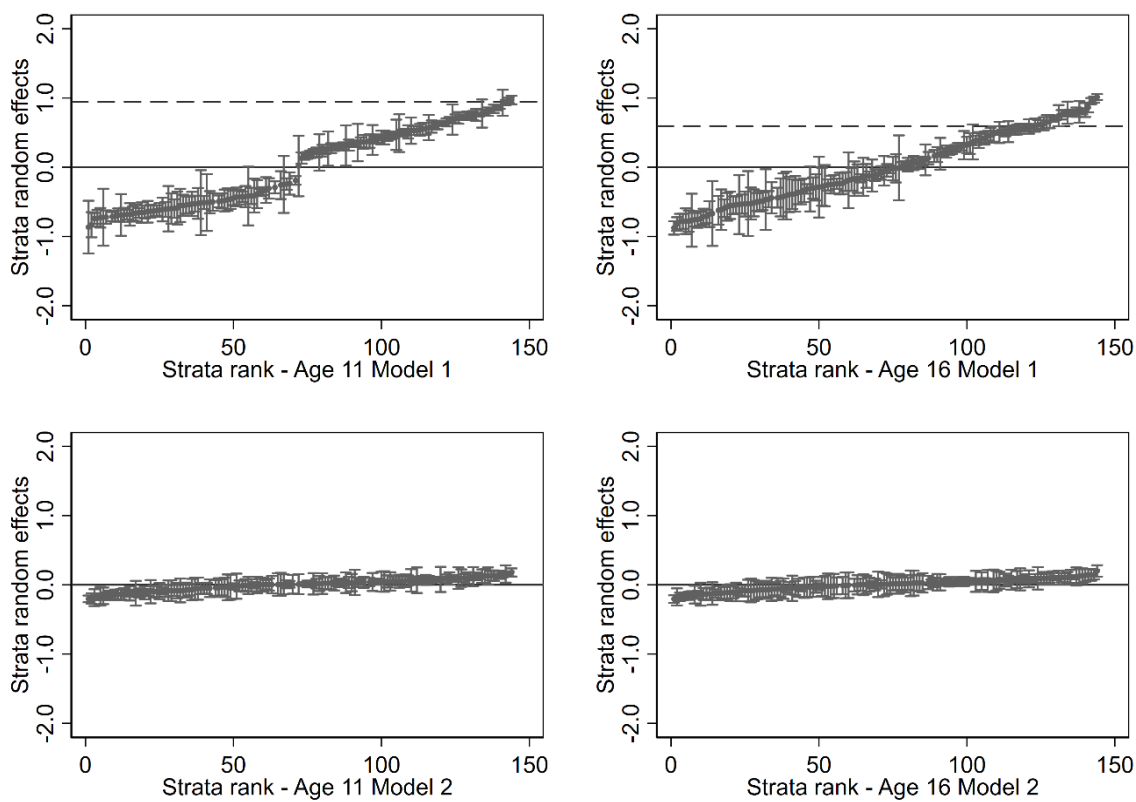

**Figure 2.**

Model 1 (unadjusted, top row) and Model 2 (adjusted, bottom row) scatterplots of predicted strata random effects (left column) and ranks (right column) for age 16 achievement (y-axis) against age 11 achievement (x-axis).

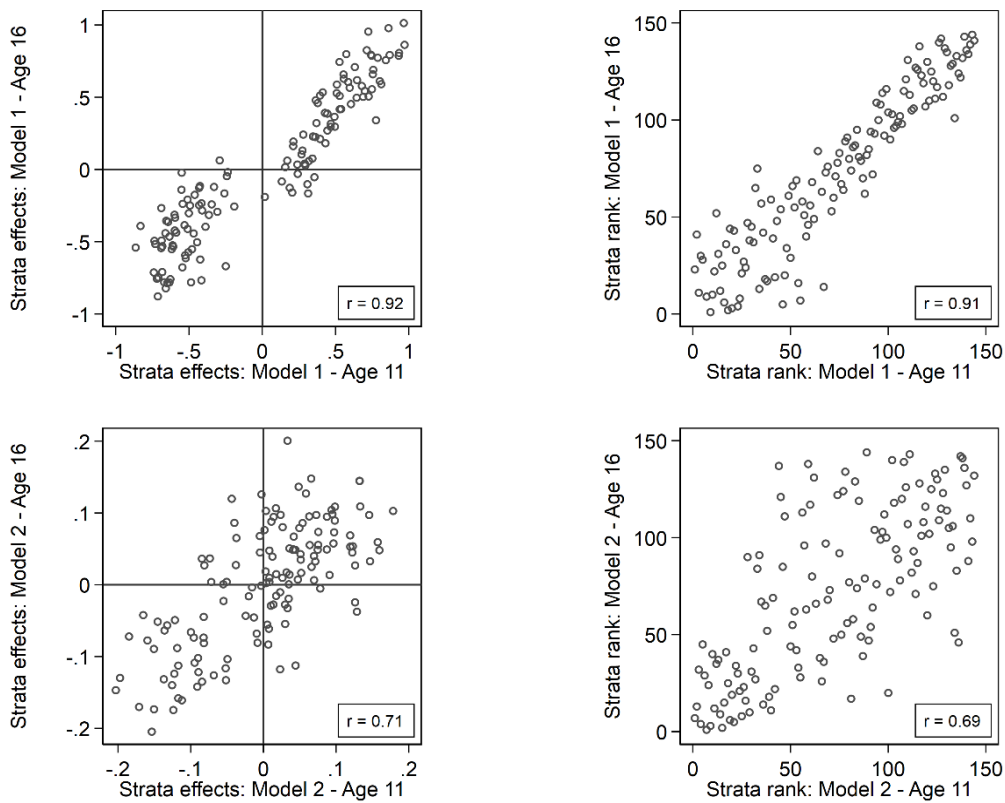

**Figure 3.**

Model 2 (adjusted) predicted strata random effects for student age 11 (left column) and age 16 (right column) achievement by intersectional sociodemographics for Black, Mixed and White ethnicities by FSM status (rows).

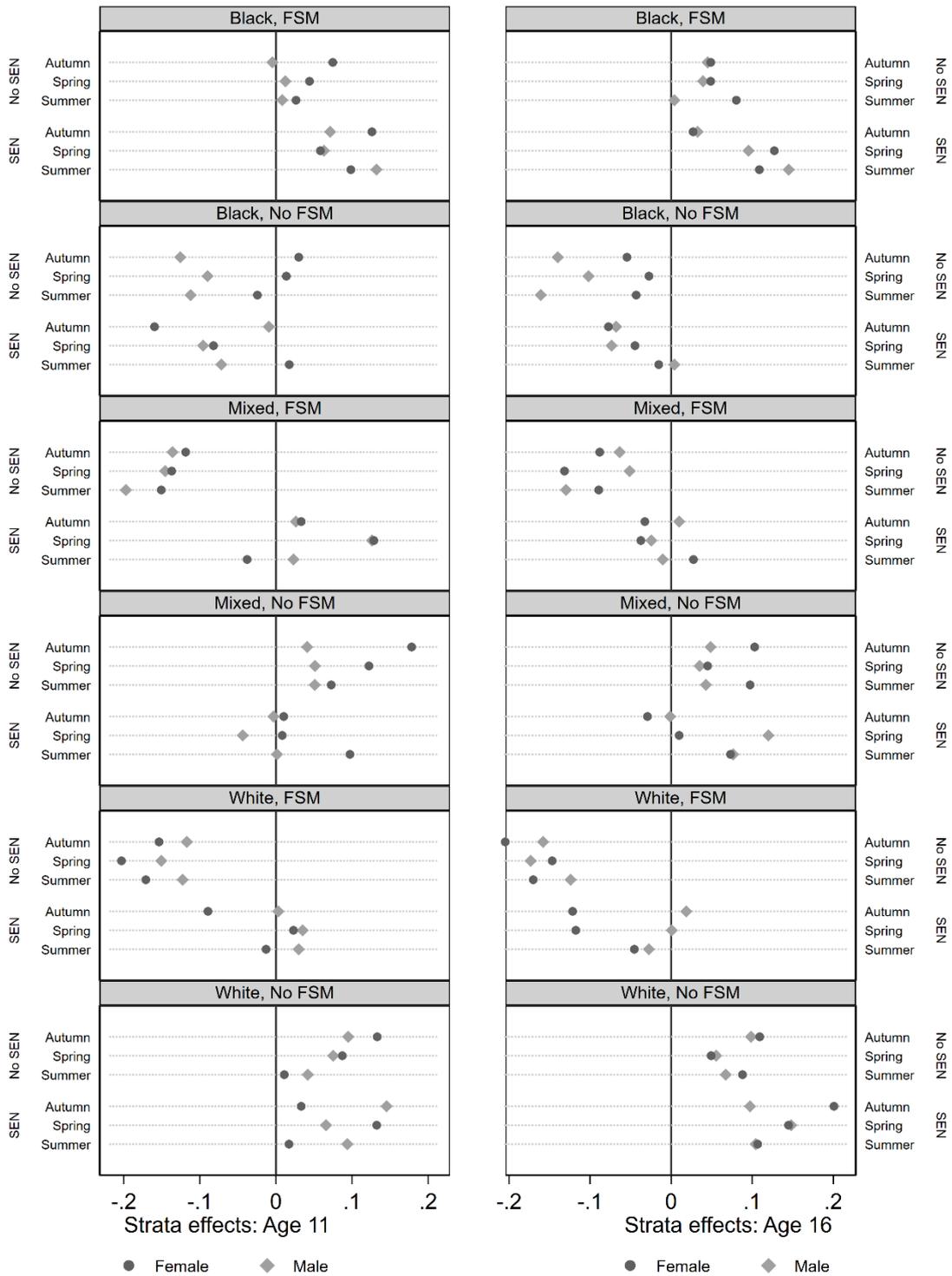

**Supplementary Material**

*Supplementary analyses*

In Supplementary Tables S2 and S3 we present the VPCs (Variance Partitioning Coefficient) and PCVs (Proportional Change in Variance) from extra models where, with all covariates entered, we additionally explore adding each two-way interaction between the sociodemographic components in turn to indicate which combination of characteristics may be most important in governing any of the remaining two- and higher-way interactional variance.

In Supplementary Tables S4 and S5 we present results from additional models where we add each component of the intersectional strata as one covariate at a time (i.e., one model for gender, one for ethnicity, and so on). These provide an assessment of the separate contribution of each component to explaining the proportion of overall variance positioned between intersectional strata.

*Supplementary reflections and limitations*

In our educational application we found intersectional strata with the fewest students coincided with the tendency towards lower achievement scores. The effect of this was seen in the intercept coefficients of our unadjusted models, which were pulled substantially below the sample mean of 0. This is consistent with the intercept representing the overall average of the stratum mean achievement scores. In contrast, a conventional linear regression model would give an intercept of 0 since the intercept under this case represents the overall average of the student achievement scores. Given this, one might ask how the presented coefficients in our adjusted model would change if a conventional linear regression were fitted. In Supplementary Table S6 we show that there are similar relationships between the multilevel and single-level linear regression models for the individual sociodemographic components. However, researchers should be aware of the potential influence of relationships between stratum size and outcome scores. Relevant to this, we note that the proportion of FSM eligible students and minority ethnicities is higher in London than for England nationally, and so imbalance would be greater if these models were fitted to other regions or the whole of England. Where researchers wish to apply these methods in other contexts, or where they are particularly interested in intersections with minority groups, similar sample restrictions or the implementation of coarseness in categories (for instance combining minor ethnicities into major ethnicity categories) may be required to successfully implement the modelling.

Our educational application also highlighted considerations researchers may have to make regarding normality assumptions for the model residuals. The student-level predicted residuals show good correspondence with a normal distribution across both outcomes and models (Supplementary Figures S5 and S7). However, the stratum-level predicted random effects show some violation of the normality assumption, particularly in Model 1 where for age 11 they demonstrate a bimodal distribution. Furthermore, some researchers may question whether it is conceptually viable to consider the intersectional identities as belonging to a wider population of identities in order to operationalise intersections as a level in the multilevel setting.

## Supplementary Tables

Table S1. Mean age 11 and age 16 achievement scores, count and percentage of the sample in each intersectional stratum. Values for strata with fewer than 10 students have been suppressed.

| Strata ID | Number of pupils | | Percentage of pupils | | Standardized score | | Term of birth | | | Gender | | FSM | | SEN | | Ethnicity | | | | | |
|---|---|---|---|---|---|---|---|---|---|---|---|---|---|---|---|---|---|---|---|---|---|
| | Age 11 | Age 16 | Age 11 | Age 16 | Age 11 | Age 16 | Aut. | Spr. | Sum. | M | F | No FSM | FSM | No SEN | SEN | White | Black | Asian | Mixed | Other | Unclass. |
| 1 | 3802 | 3050 | 4.19 | 4.28 | 0.41 | 0.22 | X | | | X | | X | | X | | X | | | | | |
| 2 | 1209 | 1084 | 1.33 | 1.52 | 0.09 | -0.07 | X | | | X | | X | | X | | | X | | | | |
| 3 | 2008 | 1616 | 2.21 | 2.27 | 0.53 | 0.44 | X | | | X | | X | | X | | | | X | | | |
| 4 | 805 | 610 | 0.89 | 0.86 | 0.40 | 0.24 | X | | | X | | X | | X | | | | | X | | |
| 5 | 393 | 324 | 0.43 | 0.45 | 0.36 | 0.29 | X | | | X | | X | | X | | | | | | X | |
| 6 | 69 | 133 | 0.08 | 0.19 | 0.40 | -0.02 | X | | | X | | X | | X | | | | | | | X |
| 7 | 758 | 487 | 0.84 | 0.68 | -0.59 | -0.63 | X | | | X | | X | | | X | X | | | | | |
| 8 | 270 | 181 | 0.30 | 0.25 | -0.85 | -0.88 | X | | | X | | X | | | X | | X | | | | |
| 9 | 261 | 171 | 0.29 | 0.24 | -0.84 | -0.62 | X | | | X | | X | | | X | | | X | | | |
| 10 | 133 | 82 | 0.15 | 0.11 | -0.71 | -0.67 | X | | | X | | X | | | X | | | | X | | |
| 11 | 57 | 33 | 0.06 | 0.05 | -0.95 | -1.04 | X | | | X | | X | | | X | | | | | X | |
| 12 | 18 | 14 | 0.02 | 0.02 | -1.38 | -1.03 | X | | | X | | X | | | X | | | | | | X |
| 13 | 1137 | 826 | 1.25 | 1.16 | -0.09 | -0.47 | X | | | X | | | X | X | | X | | | | | |
| 14 | 1279 | 1001 | 1.41 | 1.40 | -0.06 | -0.30 | X | | | X | | | X | X | | | X | | | | |
| 15 | 876 | 643 | 0.97 | 0.90 | 0.21 | 0.08 | X | | | X | | | X | X | | | | X | | | |
| 16 | 468 | 296 | 0.52 | 0.42 | -0.08 | -0.31 | X | | | X | | | X | X | | | | | X | | |
| 17 | 294 | 231 | 0.32 | 0.32 | 0.03 | 0.03 | X | | | X | | | X | X | | | | | | X | |
| 18 | 45 | 59 | 0.05 | 0.08 | -0.12 | -0.26 | X | | | X | | | X | X | | | | | | | X |
| 19 | 555 | 311 | 0.61 | 0.44 | -1.03 | -1.14 | X | | | X | | | X | | X | X | | | | | |
| 20 | 477 | 257 | 0.53 | 0.36 | -1.04 | -1.16 | X | | | X | | | X | | X | | X | | | | |
| 21 | 189 | 146 | 0.21 | 0.20 | -1.07 | -0.82 | X | | | X | | | X | | X | | | X | | | |
| 22 | 181 | 85 | 0.20 | 0.12 | -0.95 | -1.07 | X | | | X | | | X | | X | | | | X | | |
| 23 | 94 | 44 | 0.10 | 0.06 | -0.95 | -0.78 | X | | | X | | | X | | X | | | | | X | |
| 24 | 15 | SUPP | 0.02 | SUPP | -1.24 | SUPP | X | | | X | | | X | | X | | | | | | X |
| 25 | 3668 | 2698 | 4.04 | 3.78 | 0.32 | 0.13 | | X | | X | | X | | X | | X | | | | | |

| 26 | 1094 | 1056 | 1.21 | 1.48 | 0.07 | -0.08 |
|----|------|------|------|------|------|-------|
| 27 | 1961 | 1553 | 2.16 | 2.18 | 0.39 | 0.40 |
| 28 | 710 | 585 | 0.78 | 0.82 | 0.35 | 0.19 |
| 29 | 377 | 340 | 0.42 | 0.48 | 0.19 | 0.24 |
| 30 | 59 | 127 | 0.07 | 0.18 | 0.31 | 0.14 |
| 31 | 882 | 553 | 0.97 | 0.78 | -0.75 | -0.61 |
| 32 | 296 | 185 | 0.33 | 0.26 | -1.02 | -0.93 |
| 33 | 321 | 161 | 0.35 | 0.23 | -0.92 | -0.58 |
| 34 | 156 | 103 | 0.17 | 0.14 | -0.84 | -0.50 |
| 35 | 69 | 45 | 0.08 | 0.06 | -0.80 | -0.79 |
| 36 | 13 | 16 | 0.01 | 0.02 | -0.38 | -0.60 |
| 37 | 957 | 837 | 1.05 | 1.17 | -0.20 | -0.53 |
| 38 | 1185 | 965 | 1.31 | 1.35 | -0.11 | -0.34 |
| 39 | 706 | 695 | 0.78 | 0.97 | 0.13 | 0.05 |
| 40 | 424 | 314 | 0.47 | 0.44 | -0.16 | -0.34 |
| 41 | 310 | 228 | 0.34 | 0.32 | -0.03 | -0.05 |
| 42 | 38 | 65 | 0.04 | 0.09 | -0.02 | -0.43 |
| 43 | 618 | 315 | 0.68 | 0.44 | -1.06 | -1.20 |
| 44 | 482 | 301 | 0.53 | 0.42 | -1.11 | -1.13 |
| 45 | 193 | 137 | 0.21 | 0.19 | -1.15 | -0.88 |
| 46 | 244 | 88 | 0.27 | 0.12 | -0.89 | -1.18 |
| 47 | 99 | 55 | 0.11 | 0.08 | -1.03 | -0.93 |
| 48 | 14 | 18 | 0.02 | 0.03 | -0.89 | -1.27 |
| 49 | 3529 | 3063 | 3.89 | 4.29 | 0.24 | 0.13 |
| 50 | 1088 | 1024 | 1.20 | 1.44 | -0.01 | -0.16 |
| 51 | 1856 | 1633 | 2.05 | 2.29 | 0.35 | 0.42 |
| 52 | 733 | 592 | 0.81 | 0.83 | 0.30 | 0.18 |
| 53 | 402 | 345 | 0.44 | 0.48 | 0.14 | 0.05 |
| 54 | 61 | 132 | 0.07 | 0.19 | 0.01 | 0.15 |
| 55 | 1044 | 664 | 1.15 | 0.93 | -0.77 | -0.69 |
| 56 | 355 | 220 | 0.39 | 0.31 | -1.04 | -0.84 |
| 57 | 394 | 190 | 0.43 | 0.27 | -0.90 | -0.62 |
| 58 | 208 | 128 | 0.23 | 0.18 | -0.82 | -0.61 |
| 59 | 110 | 58 | 0.12 | 0.08 | -0.88 | -0.84 |
| 60 | 15 | 23 | 0.02 | 0.03 | -0.99 | -0.84 |
| 61 | 996 | 818 | 1.10 | 1.15 | -0.21 | -0.50 |
| 62 | 1187 | 998 | 1.31 | 1.40 | -0.16 | -0.40 |
| 63 | 700 | 673 | 0.77 | 0.94 | 0.04 | 0.02 |
| 64 | 451 | 315 | 0.50 | 0.44 | -0.27 | -0.46 |
| 65 | 286 | 255 | 0.32 | 0.36 | 0.03 | -0.19 |
| 66 | 32 | 47 | 0.04 | 0.07 | -0.07 | -0.55 |
| 67 | 694 | 324 | 0.76 | 0.45 | -1.12 | -1.26 |
| 68 | 606 | 333 | 0.67 | 0.47 | -1.09 | -1.09 |
| 69 | 254 | 158 | 0.28 | 0.22 | -1.02 | -0.91 |
| 70 | 292 | 126 | 0.32 | 0.18 | -1.08 | -1.17 |
| 71 | 116 | 76 | 0.13 | 0.11 | -0.89 | -0.95 |
| 72 | 15 | 11 | 0.02 | 0.02 | -0.69 | -1.23 |
| 73 | 4022 | 3059 | 4.43 | 4.29 | 0.47 | 0.42 |
| 74 | 1383 | 1197 | 1.52 | 1.68 | 0.28 | 0.21 |
| 75 | 2088 | 1666 | 2.30 | 2.34 | 0.56 | 0.64 |
| 76 | 798 | 613 | 0.88 | 0.86 | 0.57 | 0.50 |
| 77 | 410 | 301 | 0.45 | 0.42 | 0.32 | 0.46 |
| 78 | 81 | 117 | 0.09 | 0.16 | 0.56 | 0.44 |
| 79 | 387 | 372 | 0.43 | 0.52 | -0.69 | -0.31 |
| 80 | 132 | 92 | 0.15 | 0.13 | -1.07 | -0.74 |
| 81 | 124 | 104 | 0.14 | 0.15 | -0.96 | -0.39 |
| 82 | 70 | 69 | 0.08 | 0.10 | -0.67 | -0.54 |
| 83 | 25 | 26 | 0.03 | 0.04 | -1.06 | -0.72 |
| 84 | 11 | SUPP | 0.01 | SUPP | -1.02 | SUPP |
| 85 | 1312 | 982 | 1.45 | 1.38 | -0.11 | -0.33 |
| 86 | 1474 | 1195 | 1.62 | 1.68 | 0.04 | -0.10 |
| 87 | 882 | 761 | 0.97 | 1.07 | 0.25 | 0.25 |
| 88 | 600 | 414 | 0.66 | 0.58 | -0.04 | -0.14 |
| 89 | 311 | 278 | 0.34 | 0.39 | 0.11 | 0.22 |
| 90 | 59 | 71 | 0.07 | 0.10 | 0.09 | -0.04 |
| 91 | 317 | 210 | 0.35 | 0.29 | -1.13 | -1.14 |
| 92 | 220 | 180 | 0.24 | 0.25 | -0.93 | -0.97 |
| 93 | 87 | 64 | 0.10 | 0.09 | -1.26 | -0.78 |
| 94 | 114 | 74 | 0.13 | 0.10 | -0.92 | -0.97 |
| 95 | 49 | 38 | 0.05 | 0.05 | -0.98 | -0.52 |
| 96 | SUPP | SUPP | SUPP | SUPP | SUPP | SUPP |
| 97 | 3799 | 3055 | 4.19 | 4.28 | 0.36 | 0.32 |
| 98 | 1302 | 1171 | 1.43 | 1.64 | 0.20 | 0.20 |
| 99 | 2143 | 1600 | 2.36 | 2.24 | 0.46 | 0.61 |
| 100 | 840 | 653 | 0.93 | 0.92 | 0.44 | 0.39 |
| 101 | 408 | 342 | 0.45 | 0.48 | 0.16 | 0.29 |
| 102 | 86 | 126 | 0.09 | 0.18 | 0.19 | 0.45 |
| 103 | 433 | 383 | 0.48 | 0.54 | -0.65 | -0.42 |
| 104 | 163 | 103 | 0.18 | 0.14 | -1.01 | -0.71 |
| 105 | 147 | 111 | 0.16 | 0.16 | -0.87 | -0.55 |
| 106 | 83 | 55 | 0.09 | 0.08 | -0.74 | -0.50 |
| 107 | 30 | 32 | 0.03 | 0.04 | -1.15 | -0.66 |
| 108 | SUPP | 11 | SUPP | 0.02 | SUPP | -0.38 |
| 109 | 1207 | 941 | 1.33 | 1.32 | -0.23 | -0.31 |
| 110 | 1444 | 1152 | 1.59 | 1.62 | -0.06 | -0.14 |
| 111 | 782 | 782 | 0.86 | 1.10 | 0.13 | 0.14 |
| 112 | 560 | 391 | 0.62 | 0.55 | -0.13 | -0.24 |
| 113 | 313 | 282 | 0.34 | 0.40 | -0.02 | 0.09 |
| 114 | 40 | 56 | 0.04 | 0.08 | -0.18 | -0.17 |
| 115 | 340 | 242 | 0.37 | 0.34 | -1.05 | -1.17 |
| 116 | 287 | 178 | 0.32 | 0.25 | -1.10 | -0.87 |
| 117 | 108 | 92 | 0.12 | 0.13 | -1.10 | -0.92 |
| 118 | 125 | 69 | 0.14 | 0.10 | -0.83 | -1.02 |
| 119 | 47 | 38 | 0.05 | 0.05 | -0.83 | -0.68 |
| 120 | SUPP | 11 | SUPP | 0.02 | SUPP | -1.04 |
| 121 | 4052 | 3201 | 4.47 | 4.49 | 0.23 | 0.34 |
| 122 | 1307 | 1251 | 1.44 | 1.75 | 0.11 | 0.16 |
| 123 | 2089 | 1553 | 2.30 | 2.18 | 0.32 | 0.58 |
| 124 | 822 | 672 | 0.91 | 0.94 | 0.34 | 0.43 |
| 125 | 397 | 347 | 0.44 | 0.49 | 0.16 | 0.26 |
| 126 | 79 | 129 | 0.09 | 0.18 | 0.14 | 0.39 |
| 127 | 537 | 429 | 0.59 | 0.60 | -0.83 | -0.49 |
| 128 | 182 | 126 | 0.20 | 0.18 | -0.91 | -0.68 |
| 129 | 199 | 124 | 0.22 | 0.17 | -0.95 | -0.61 |
| 130 | 111 | 72 | 0.12 | 0.10 | -0.64 | -0.39 |
| 131 | 50 | 38 | 0.06 | 0.05 | -1.02 | -0.84 |
| 132 | SUPP | 13 | SUPP | 0.02 | SUPP | -0.73 |
| 133 | 1237 | 1009 | 1.36 | 1.41 | -0.25 | -0.36 |
| 134 | 1492 | 1187 | 1.64 | 1.66 | -0.12 | -0.13 |
| 135 | 802 | 701 | 0.88 | 0.98 | 0.01 | 0.16 |
| 136 | 565 | 414 | 0.62 | 0.58 | -0.19 | -0.21 |
| 137 | 303 | 244 | 0.33 | 0.34 | -0.03 | 0.11 |
| 138 | 37 | 45 | 0.04 | 0.06 | 0.12 | 0.02 |
| 139 | 456 | 241 | 0.50 | 0.34 | -1.15 | -1.09 |
| 140 | 339 | 196 | 0.37 | 0.27 | -1.10 | -0.92 |
| 141 | 118 | 100 | 0.13 | 0.14 | -1.06 | -0.74 |
| 142 | 143 | 78 | 0.16 | 0.11 | -1.14 | -0.91 |
| 143 | 59 | 25 | 0.07 | 0.04 | -1.02 | -0.84 |
| 144 | SUPP | 13 | SUPP | 0.02 | SUPP | -0.95 |

Note:

Aut. = Autumn, Spr. = Spring, Sum. = Summer, M = Male, F = Female, Unclass. = Unclassified.

Table S2. Proportional Change Variance (PCV) and Variance Partitioning Coefficient (VPC) from models predicting age 11 standardized scores, adjusting for stratum sociodemographic components and each possible two-way interaction between these in turn.

| Interaction | VPC | PCV |
|---|---|---|
| Age*Gender | 1.33% | 0.50% |
| Age*FSM | 1.32% | 1.30% |
| Age*SEN | 1.27% | 4.47% |
| Age*Ethnicity | 1.32% | 1.11% |
| Gender*FSM | 1.32% | 0.90% |
| Gender*SEN | 1.32% | 1.20% |
| Gender*Ethnicity | 1.29% | 3.47% |
| FSM*SEN | 1.14% | 14.63% |
| FSM*Ethnicity | 0.74% | 44.81% |
| SEN*Ethnicity | 0.93% | 30.39% |

Table S3. Proportional Change Variance (PCV) and Variance Partitioning Coefficient (VPC) from models predicting age 16 standardized scores, adjusting for stratum sociodemographic components and each possible two-way interaction between these in turn.

| Interaction | VPC | PCV |
|---|---|---|
| Age*Gender | 1.24% | 0.87% |
| Age*FSM | 1.25% | 0.06% |
| Age*SEN | 1.24% | 0.78% |
| Age*Ethnicity | 1.20% | 4.45% |
| Gender*FSM | 1.24% | 1.38% |
| Gender*SEN | 1.24% | 1.35% |
| Gender*Ethnicity | 1.20% | 4.29% |
| FSM*SEN | 1.24% | 1.08% |
| FSM*Ethnicity | 0.19% | 85.08% |
| SEN*Ethnicity | 1.01% | 19.67% |

Table S4. Results predicting age 11 standardized scores in multilevel models, adjusting for each stratum sociodemographic component in turn.

| | | Est. | SE | Est. | SE | Est. | SE | Est. | SE | Est. | SE |
|---|---|---|---|---|---|---|---|---|---|---|---|
| Intercept | | -0.359 | 0.083 | -0.411 | 0.067 | -0.278 | 0.066 | 0.128 | 0.023 | -0.411 | 0.115 |
| Term | Autumn | | | | | | | | | | |
| | Spring | -0.043 | 0.117 | | | | | | | | |
| | Summer | -0.085 | 0.117 | | | | | | | | |
| Gender | Male | | | | | | | | | | |
| | Female | | | 0.019 | 0.096 | | | | | | |
| FSM | No FSM | | | | | | | | | | |
| | FSM | | | | | -0.246 | 0.093 | | | | |
| SEN | No SEN | | | | | | | | | | |
| | SEN | | | | | | | -1.076 | 0.034 | | |
| Ethnicity | White | | | | | | | | | | |
| | Black | | | | | | | | | -0.087 | 0.163 |
| | Asian | | | | | | | | | 0.052 | 0.163 |
| | Mixed | | | | | | | | | 0.051 | 0.163 |
| | Other | | | | | | | | | -0.002 | 0.164 |
| | Unclassified | | | | | | | | | 0.045 | 0.168 |
| | | Est. | SE | Est. | SE | Est. | SE | Est. | SE | Est. | SE |
| Strata var. | | 0.319 | 0.039 | 0.32 | 0.039 | 0.304 | 0.037 | 0.035 | 0.005 | 0.317 | 0.038 |
| Student var. | | 0.766 | 0.004 | 0.766 | 0.004 | 0.766 | 0.004 | 0.765 | 0.004 | 0.766 | 0.004 |
| VPC | | 29.4% | | 29.5% | | 28.4% | | 4.4% | | 29.3% | |
| PCV | | 0.4% | | 0.1% | | 5.0% | | 88.9% | | 0.9% | |

Note:

Est. = estimate, SE = standard error, Term = Term of Birth, Unclass. = Unclassified, var. = variance.

Table S5. Results predicting age 16 standardized scores in multilevel models, adjusting for each stratum sociodemographic component in turn.

| | | Est. | SE | Est. | SE | Est. | SE | Est. | SE | Est. | SE |
|---|---|---|---|---|---|---|---|---|---|---|---|
| Intercept | | -0.339 | 0.075 | -0.471 | 0.06 | -0.17 | 0.057 | 0.057 | 0.032 | -0.46 | 0.102 |
| Term | Autumn | | | | | | | | | | |
| | Spring | -0.036 | 0.106 | | | | | | | | |
| | Summer | -0.062 | 0.106 | | | | | | | | |
| Gender | Male | | | | | | | | | | |
| | Female | | | 0.201 | 0.085 | | | | | | |
| FSM | No FSM | | | | | | | | | | |
| | FSM | | | | | -0.401 | 0.08 | | | | |
| SEN | No SEN | | | | | | | | | | |
| | SEN | | | | | | | -0.877 | 0.048 | | |
| Ethnicity | White | | | | | | | | | | |
| | Black | | | | | | | | | -0.041 | 0.145 |
| | Asian | | | | | | | | | 0.272 | 0.145 |
| | Mixed | | | | | | | | | 0.078 | 0.145 |
| | Other | | | | | | | | | 0.152 | 0.146 |
| | Unclassified | | | | | | | | | 0.071 | 0.15 |
| | | Est. | SE | Est. | SE | Est. | SE | Est. | SE | Est. | SE |
| Strata var. | | 0.261 | 0.032 | 0.251 | 0.031 | 0.221 | 0.027 | 0.073 | 0.009 | 0.073 | 0.009 |
| Student var. | | 0.804 | 0.004 | 0.804 | 0.004 | 0.804 | 0.004 | 0.804 | 0.004 | 0.804 | 0.004 |
| VPC | | 24.5% | | 23.8% | | 21.6% | | 8.3% | | 23.7% | |
| PCV | | 0.3% | | 3.9% | | 15.5% | | 72.2% | | 4.3% | |

Note:

Est. = estimate, SE = standard error, Term = Term of Birth, Unclass. = Unclassified, var. = variance.

Table S6. Results predicting age 11 and age 16 standardized scores in single-level models unadjusted (Model 1) and adjusted (Model 2) for stratum component sociodemographic main effects.

| | | Model 1: Unadjusted | | | | Model 2. Adjusted for main effects | | | |
| | | Age 11 score | | Age 16 score | | Age 11 score | | Age 16 score | |
| | | Est. | SE | Est. | SE | Est. | SE | Est. | SE |
|---|---|---|---|---|---|---|---|---|---|
| | Intercept | 0.000 | 0.003 | 0.000 | 0.004 | 0.374 | 0.007 | 0.158 | 0.008 |
| Term | Autumn | | | | | | | | |
| | Spring | | | | | -0.090 | 0.007 | -0.053 | 0.008 |
| | Summer | | | | | -0.161 | 0.007 | -0.068 | 0.008 |
| Gender | Male | | | | | | | | |
| | Female | | | | | 0.030 | 0.006 | 0.199 | 0.007 |
| FSM | No FSM | | | | | | | | |
| | FSM | | | | | -0.348 | 0.006 | -0.465 | 0.007 |
| SEN | No SEN | | | | | | | | |
| | SEN | | | | | -1.051 | 0.008 | -0.837 | 0.010 |
| Ethnicity | White | | | | | | | | |
| | Black | | | | | -0.094 | 0.008 | -0.073 | 0.009 |
| | Asian | | | | | 0.114 | 0.008 | 0.301 | 0.009 |
| | Mixed | | | | | 0.031 | 0.010 | 0.054 | 0.012 |
| | Other | | | | | -0.018 | 0.013 | 0.141 | 0.015 |
| | Unclass. | | | | | -0.024 | 0.031 | 0.021 | 0.026 |
| | | Est. | SE | Est. | SE | Est. | SE | Est. | SE |
| | Residual | 1.000 | 0.005 | 1.000 | 0.005 | 0.773 | 0.004 | 0.812 | 0.004 |

Note:

Est. = estimate, SE =  standard error, Term  = Term of birth,  Unclass. = Unclassified.

*Supplementary Figures*

Figure S1. Histograms of non-standardized (left) and standardized and normalized (right) achievement scores for age 11 (top row) and age 16 (bottom row).

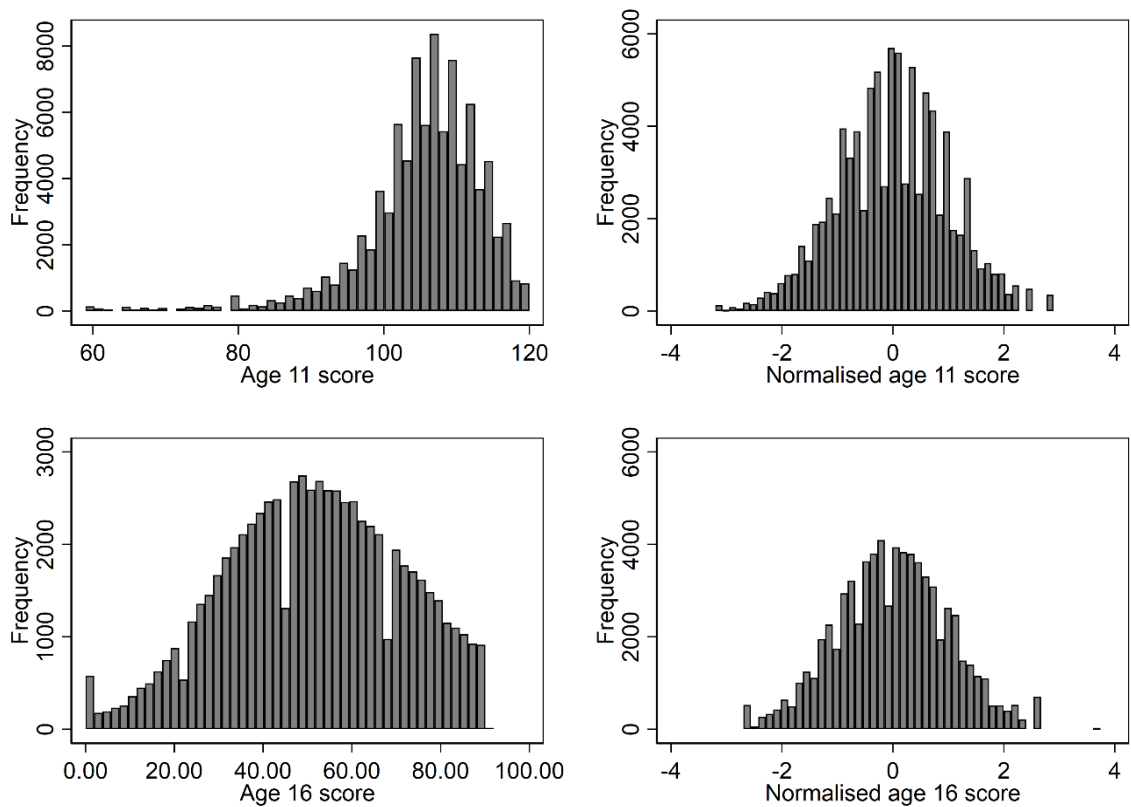

Figure S2. Average achievement scores by stratum size.

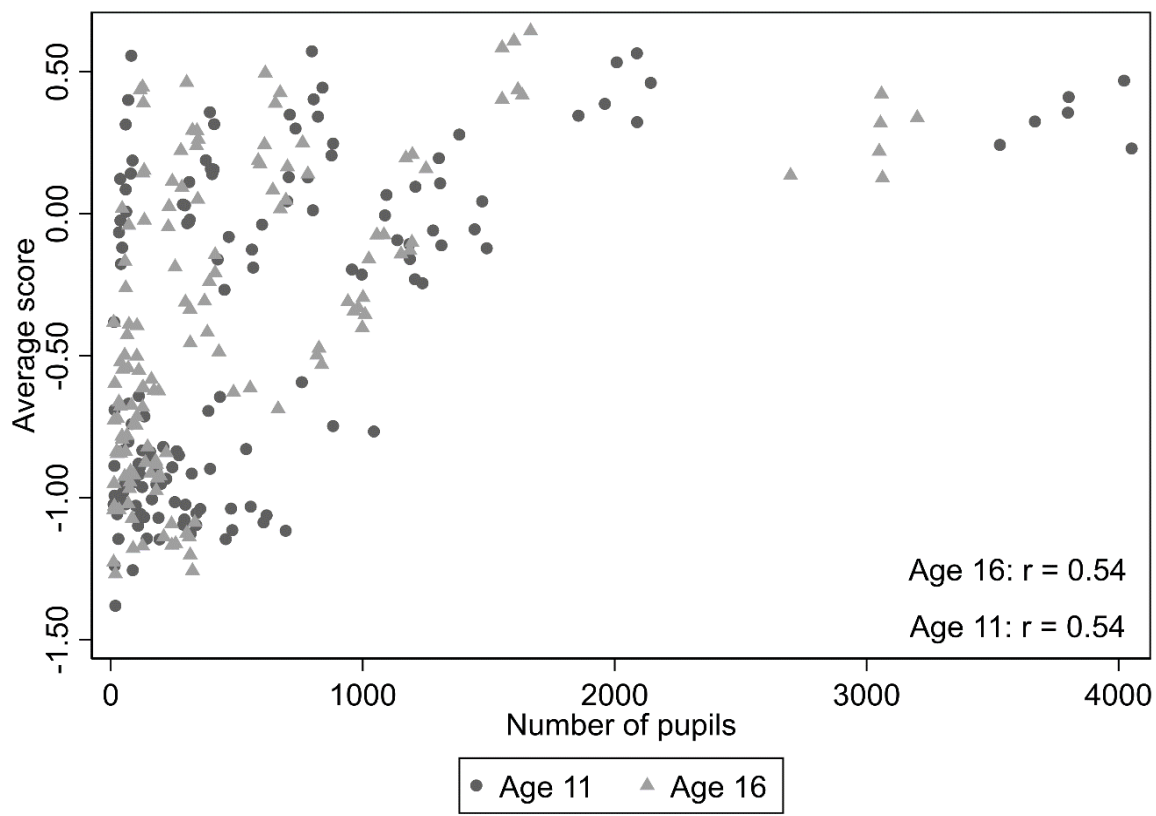

Figure S3. Descriptive versus multilevel stratum mean scores (from Model 1) for age 11 (left) and age 16 (right) standardized achievement scores.

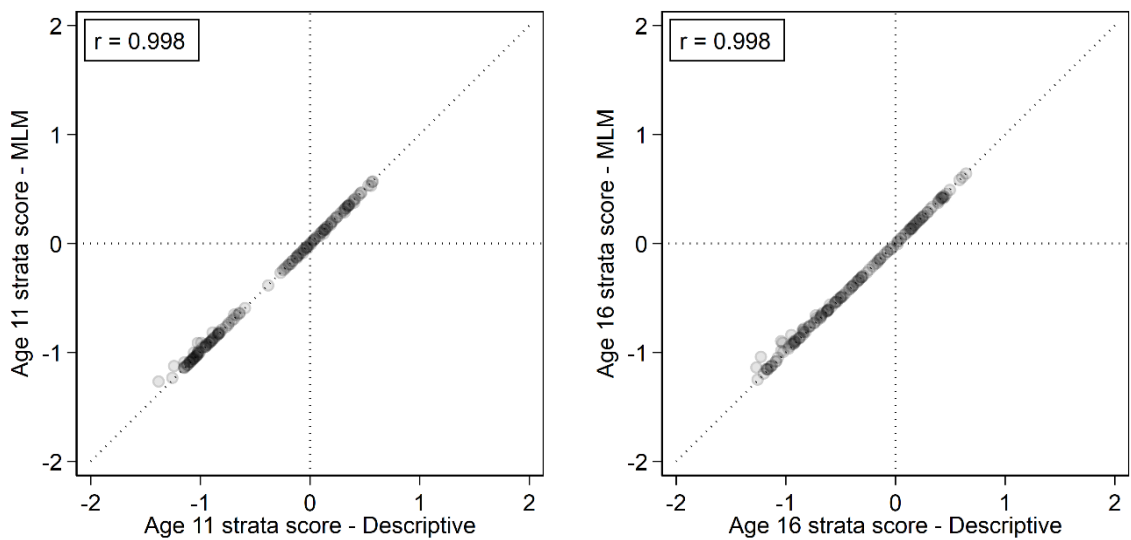

Figure S4. Difference between shrunken and un-shrunken stratum residuals for Model 1 (unadjusted) by stratum size for age 11 (left) and age 16 (right).

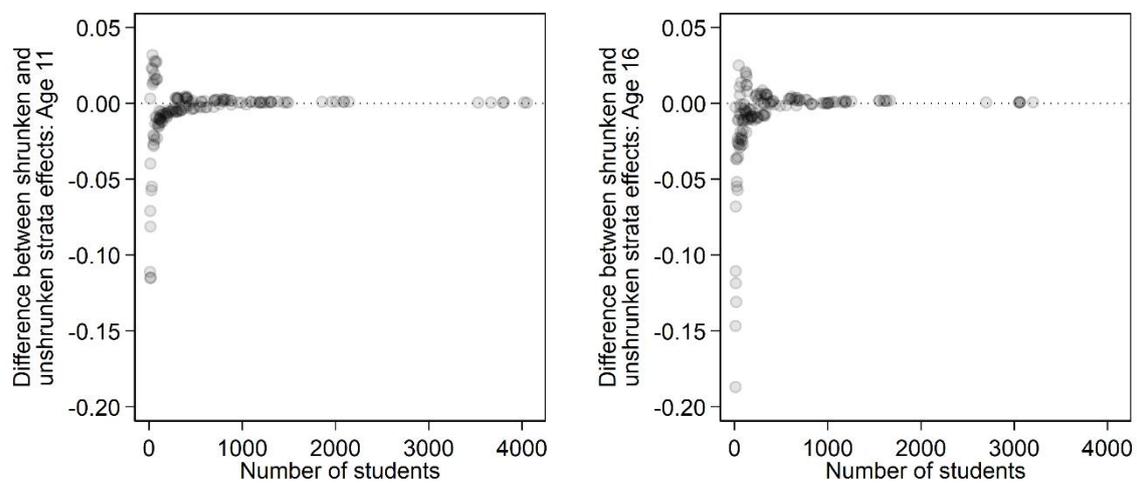

Figure S5. Student (left) and stratum (right) level residuals for age 11 achievement for Model 1 (unadjusted, top row) and Model 2 (adjusted for main effects, bottom row).

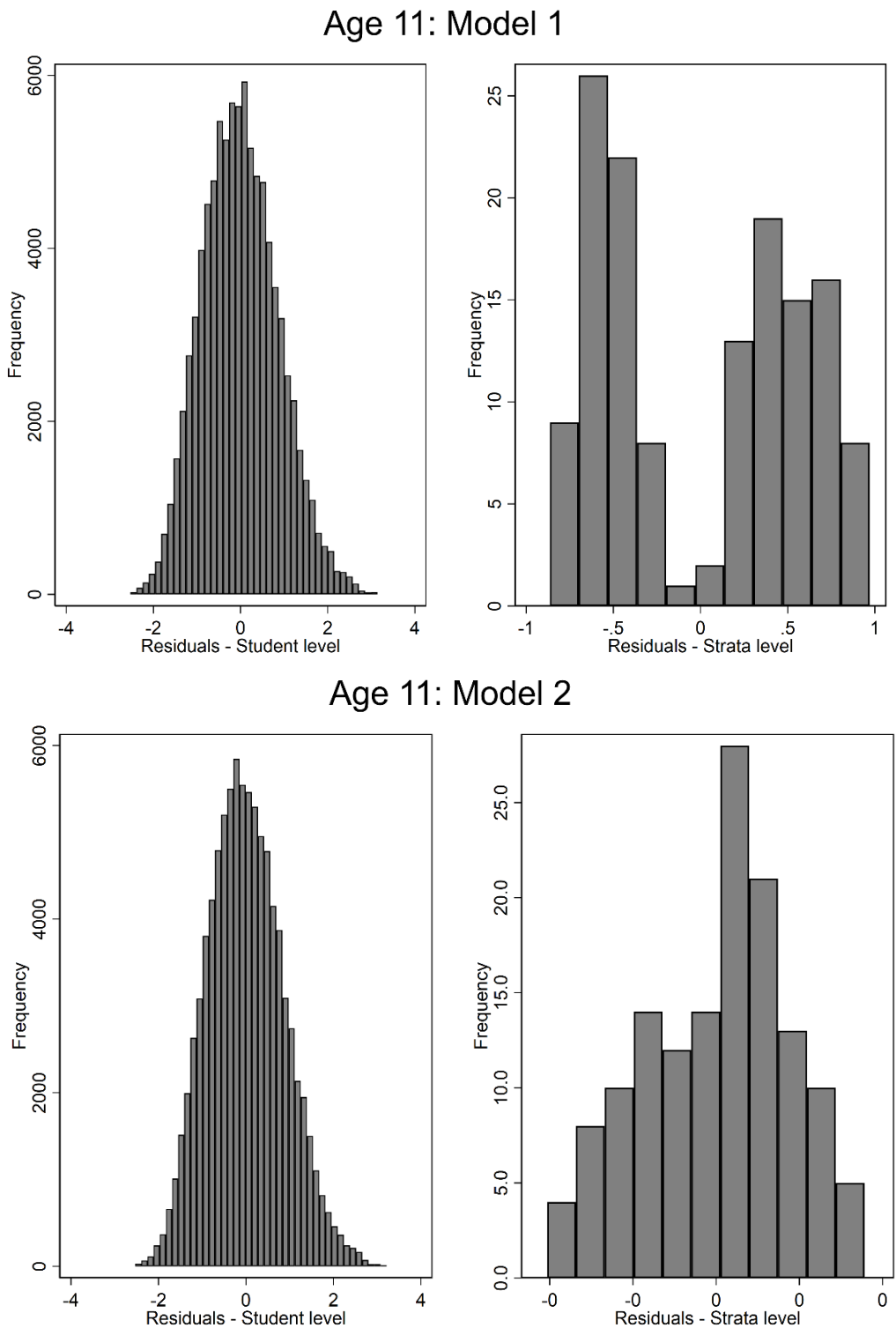

Figure S6. Change in stratum effects between age 11 and age 16 cohorts for Model 1 (unadjusted, left) and Model 2 (adjusted for main effects, right).

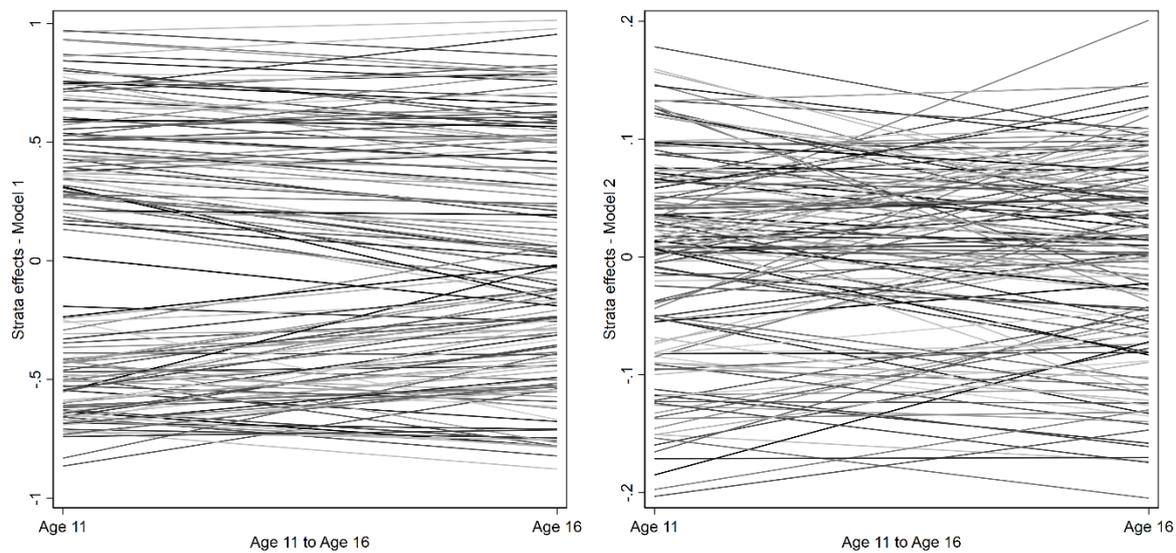

Figure S7. Student (left) and stratum (right) level residuals for age 16 achievement for Model 1 (unadjusted, top row) and Model 2 (adjusted for main effects, bottom row).

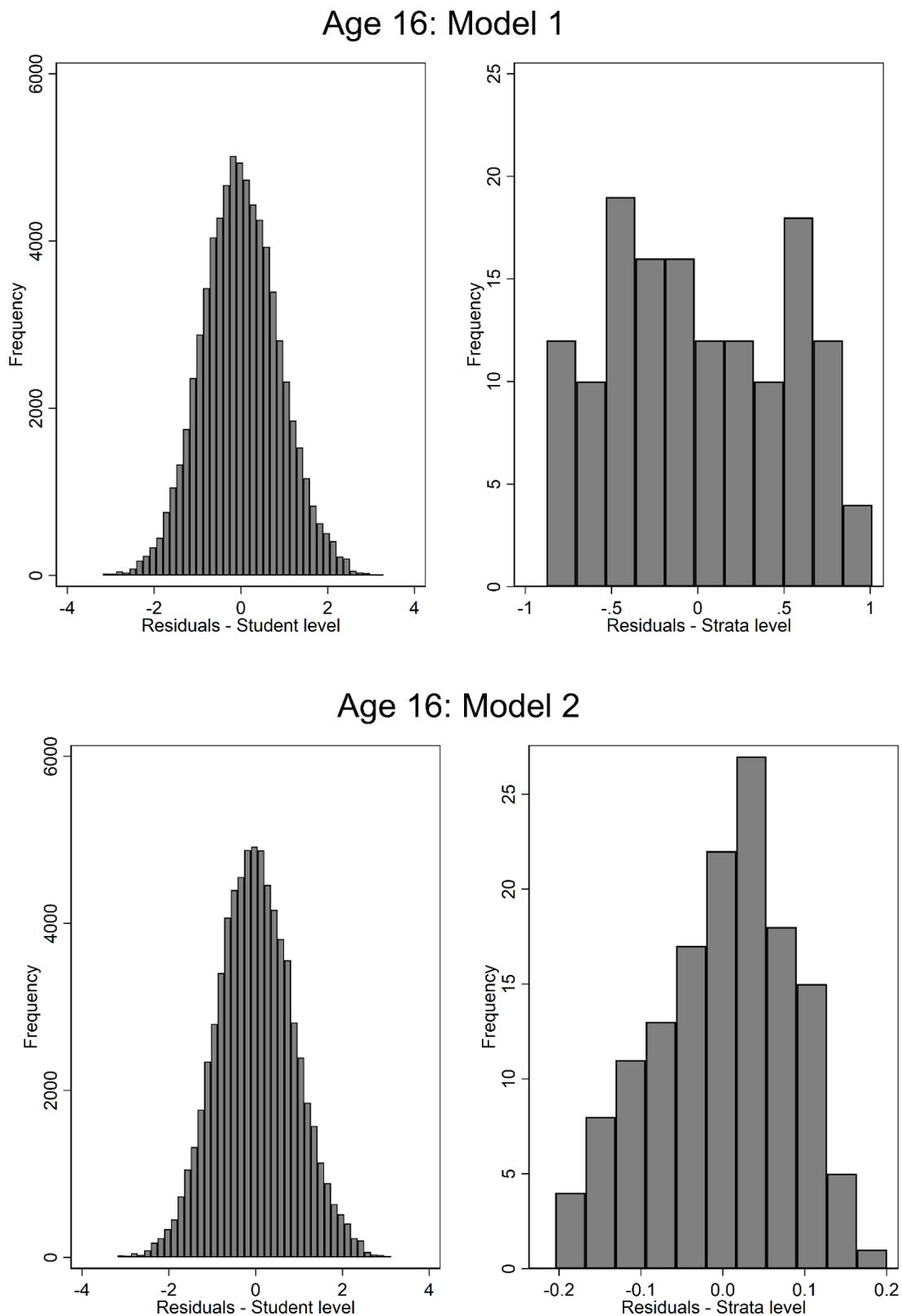

*Supplementary Stata code*

```
***************************************************************************
* Intersectional MAIHDA
***************************************************************************
***************************************************************************
* Descriptive statistics
***************************************************************************
* Load the data
use "data.dta", clear

* Summarize the continuous outcome
summarize y

* Tabulate the categorical covariates
tab1 x1 x2 x3 x4 x5 x6

* Generate the statum ID
egen stratum = group(x1 x2 x3 x4 x5 x6)
order stratum

* Generate the number of students per stratum
bysort stratum: generate n = _N

* Generate the student ID
generate student = _n
order student, after(stratum)

* List the number of strata
codebook stratum

* Tabulate the number of students per stratum
tabulate stratum

* Calculate the observed stratum means
```

```
bysort stratum: egen ymean = mean(y)
format %3.1f ymean

* Tag one student per stratum
egen stratumtag = tag(stratum)

****************************************************************************
* Model 1: Unadjusted model
****************************************************************************
* Fit model
mixed y || stratum:

* Store model results
estimates store model1

* Calculate the VPC
estat icc

* Store stratum variance as a scalar
scalar m1sigma2u = exp(_b[lns1_1_1:_cons])^2
scalar list m1sigma2u

* Predict the stratum means
predict m1ymn, fitted

* Predict the stratum random effects values
predict m1u, reffects

* Predict the stratum random effects standard errors
predict m1use, reses

* Predict the stratum means standard errors
generate m1ymnse = m1use
```

```
* Calculate the lower and upper bounds of the predicted stratum means 95%
CIs

generate m1ymnlo = m1ymn -1.96*m1use

generate m1ymnhi = m1ymn +1.96*m1use

* Rank the predicted stratum means

egen m1ymnrank = rank(m1ymn) if stratumtag == 1

bysort stratum (m1ymnrank): replace m1ymnrank = m1ymnrank[1]

****************************************************************************
* Model 2: Adjusted model
****************************************************************************
* Fit model

mixed y i.x1 i.x2 i.x3 i.x4 i.x5 i.x6 || stratum:

* Store model results

estimates store model2

* Calculate the VPC

estat icc

* Store stratum variance as a scalar

scalar m2sigma2u = exp(_b[lns1_1_1:_cons])^2

scalar list m1sigma2u

* Calculate PCV

display "PCV = " %3.1f 100*((m1sigma2u - m2sigma2u)/m1sigma2u) "%"

* Predict the stratum means

predict m2ymn, fitted

* Rank the predicted stratum means

egen m2ymnrank = rank(m2ymn) if stratumtag == 1

bysort stratum (m2ymnrank): replace m2ymnrank = m2ymnrank[1]
```

```
* Predict the stratum random effects values
predict m2u, reffects

* Predict the stratum random effects standard errors
predict m2use, reses

* Calculate the stratum random effects 95% CI lower and upper bounds
generate m2ulo = m2u -1.96*m2use
generate m2uhi = m2u +1.96*m2use

* Rank the predicted stratum random effects values
egen m2urank = rank(m2u) if stratumtag == 1
bysort stratum (m2urank): replace m2urank = m2urank[1]

******************************************************************************
* Stratum level table of results
******************************************************************************
* Keep variables of interest
keep stratum n x1 x2 x3 x4 x5 ///
      ymean ///
      m1ymn m1ymnrank m1ymnse m1ymnlo m1ymnhi ///
      m2u m2use m2ulo m2uhi m2urank

* Reduce the data to one row per stratum
duplicates drop

* List the data
format %3.1f ymean ///
      m1ymn m1ymnse m1ymnlo m1ymnhi ///
      m2u m2use m2ulo m2uhi
list

* Model 1 caterpillar plot of predicted stratum means
serrbar m1ymn m1ymnse m1ymnrank, scale(1.96)
```

```
* Model 2 caterpillar plot of predicted stratum random effects
serrbar m2u m2use m2urank, scale(1.96) yline(0)

****************************************************************************
exit
```